\documentclass[twocol]{ametsocV6.1}

\usepackage{placeins}
\usepackage{xcolor}
\usepackage{outlines}
\title{Controls on the ocean response to idealized Antarctic meltwater input}

\authors{Rory Basinski-Ferris\aff{a,b}, Laure Zanna\aff{a}, Ian Eisenman\aff{b} \correspondingauthor{Rory Basinski-Ferris, rbasinskiferris@ucsd.edu}}

\affiliation{\aff{a}{Courant Institute School of Mathematics, Computing, and Data Science, New York University, New York, NY, USA} \\ \aff{b}{Scripps Institution of Oceanography, University of California San Diego, La Jolla, CA, USA}}

\abstract{Antarctic meltwater is expected to increase throughout the coming centuries and impact sea level, ocean circulation, and the coupled climate evolution. This motivates interest in understanding the ocean response to Antarctic freshwater injection, including potential sources of uncertainty. In this study, we use idealized single-basin ocean simulations with meltwater input to examine the dependence of ocean transport and the timescales of the adjustment of regional sea level patterns on: (a) the model resolution and parameter values such as the mesoscale eddy Gent-McWilliams parameterization and vertical diffusivity, thereby partially addressing structural and parametric uncertainty; and (b) the depth of meltwater forcing, which must be prescribed both in our experiments and in most comprehensive climate model simulations, due to a lack of dynamic coupling with an ice sheet model. We find distinct sea level adjustment timescales and changes in the upper and abyssal cells depending on the depth of input, including a near total shutdown of the abyssal cell which only occurs with meltwater injection at the surface. We additionally find correlations between the ocean response to meltwater and the background stratification in each control simulation, which depends on the model resolution and parameter values. These results indicate that, in addition to uncertainty in how ocean models interact with fluxes from ice sheets, the ocean physics and simulated preindustrial state substantially influence the dynamic ocean response to projected ice shelf meltwater fluxes. }

\begin{document}

%%%%%%%%%%%%%%%%%%%%%%%%%%%%%%%%%%%%%%%%%%%%%%%%%%%%%%%%%%%%%%%%%%%%%
% SIGNIFICANCE STATEMENT/CAPSULE SUMMARY
%%%%%%%%%%%%%%%%%%%%%%%%%%%%%%%%%%%%%%%%%%%%%%%%%%%%%%%%%%%%%%%%%%%%%
%
% If you are including an optional significance statement for a journal article or a required capsule summary for BAMS 
% (see www.ametsoc.org/ams/index.cfm/publications/authors/journal-and-bams-authors/formatting-and-manuscript-components for details), 
% please apply the necessary command as shown below:
%
% Significance Statement (all journals except BAMS)
%

%% Necessary!
\maketitle

\statement
The amount of meltwater that Antarctica releases into the ocean is expected to increase in the future. This study looks at uncertainties in how the ocean responds to Antarctic meltwater using an idealized single-basin ocean model. We show that the ocean model resolution and parameter values, as well as the depth of meltwater injection, affect the simulated ocean response. These results indicate potential sources of uncertainty in projections of ocean circulation and sea level changes due to Antarctic meltwater.

\section{Introduction}
Antarctica has been losing mass over recent decades \citep[e.g.,][]{Otosaka2023Mass2020}, with projections indicating continued mass loss during the coming centuries \citep[e.g.,][]{Oppenheimer2019SeaCommunities}.  This meltwater input from the Antarctic ice sheet raises sea level, with an inhomogeneous pattern determined by changes in gravitation, rotation, and solid-earth deformation \citep[e.g.,][]{Farrell1976OnLevel,Kopp2010TheExperiments,Mitrovica2018QuantifyingFlux}, along with the dynamic redistribution of heat and salinity by the ocean circulation \citep[e.g.,][]{Stammer2008ResponseMelting,Lorbacher2012RapidMelting,Kopp2010TheExperiments,Schmidt2023AnomalousForcing}. The introduction of Antarctic meltwater also affects the ocean state more broadly. It has been demonstrated to change the large-scale ocean circulation including the abyssal cell \citep[e.g., ][]{Lago2019ProjectedContributions,Li2023AbyssalMeltwater,Moon2025AntarcticCirculation}, change the ocean stratification, and lead to Southern Ocean surface cooling and subsurface warming \citep[e.g.,][]{Bronselaer2018ChangeMeltwater, Schmidt2023AnomalousForcing, Li2023GlobalStudy}, amongst other impacts. Modulated by these effects on the ocean state, meltwater introduction is expected to additionally change the coupled climate response to anthropogenic forcing, including standard metrics such as the global mean temperature change \citep[e.g.,][]{Bronselaer2018ChangeMeltwater, Sadai2020FutureWarming, Li2023GlobalStudy}, through influencing both ocean heat uptake and radiative feedbacks via the pattern effect \citep{Dong2022AntarcticEffect}. Thus, understanding the sensitivities of the modeled ocean dynamic response to Antarctic meltwater input is key for a wide range of climate change observables.\\

Simulated features of the ocean circulation, including large-scale systems such as the meridional overturning circulation and Antarctic Circumpolar Current, are sensitive to the grid resolution \citep[e.g.,][]{Roberts2020SensitivityChanges,Marques2022NeverWorld2:Resolutions} and choice of parameter values including vertical diffusivity \citep[e.g.,][]{Mignot2006ADiffusivity} and eddy diffusivities \citep[e.g.,][]{Marshall2017TheStudy,Saenko2018ImpactModel}. This sensitivity affects both the preindustrial control state and the response of the ocean to forcing (e.g., changes in the atmospheric CO$_2$ concentration) including projected changes in transport, heat uptake, and dynamic sea level \citep[e.g.,][]{Saenko2018ImpactModel,Huber2017DriversUptake,Todd2020OceanOnlyChange,Wickramage2023SensitivityResolution}. While there have been explorations of these sensitivities for the atmosphere-ocean coupled system forced by anthropogenic emissions, they are yet to be explored when including interactions with ice sheets. Similar to the response to atmospheric fluxes, the impact of meltwater fluxes on the ocean state may also be expected to be sensitive to resolution and model parameter values, due to anticipated dependencies on the represented stratification and circulation \citep{Swart2023TheDesign}. This sensitivity will introduce uncertainty in projections of the response to Antarctic meltwater stemming from the ocean model set-up alone, which has not previously been systematically explored.
\\

The introduction of meltwater into the model ocean is associated with its own uncertainty, partially because most current coupled climate models, including those in phases 5 and 6 of the Coupled Model Intercomparison Project (CMIP), have fixed ice sheets which do not interact with the ocean \citep{Taylor2012AnDesign,Eyring2016OverviewOrganization}. Thus,  the basal melt of ice shelves, which is a dominant source of mass loss \citep[e.g.,][]{Pritchard2012AntarcticShelves,Rignot2013Ice-shelfAntarctica,Depoorter2013CalvingShelves}, is not dynamically represented. Due to this lack of coupling, choices must be made regarding the magnitude and spatiotemporal distribution of prescribed meltwater input, with current community efforts aimed at investigating the sensitivity to these choices \citep{Swart2023TheDesign}. For example, one such choice that needs to be made is the depth of meltwater prescription, which has typically been chosen to be at the ocean surface \citep[e.g.,][]{Stammer2008ResponseMelting,Lorbacher2012RapidMelting,Kopp2010TheExperiments,Bronselaer2018ChangeMeltwater,Golledge2019GlobalMelt,Moorman2020ThermalModel,Li2023GlobalStudy,Schmidt2023AnomalousForcing,Park2023FutureModel}, whereas observational evidence implies that much of the flux occurs considerably below the surface \citep{Kim2016TheNeon,Garabato2017VigorousShelf}, potentially dependent on the local stratification \citep{Arnscheidt2021OnShelves}. \cite{Eisenman2024TheFluxes} and \cite{Basinski-Ferris2025AEvolution} recently demonstrated that the depth of the meltwater flux has a substantial impact on the ocean dynamical response and sea level. \\

Here, we investigate how the ocean response to meltwater depends on modelling choices, including the ocean model parameter values. In particular, we investigate the impact of different ocean states on the global adjustment of the ocean to both surface and subsurface idealized meltwater input to highlight physical dependencies of the response. We use idealized meltwater input and a simplified model set-up to systematically explore a wide range of ocean model parameter perturbations and assess their impact on the ocean's response. This follows previous literature on perturbed parameter ensembles where parameters are varied through a range of values to sample model uncertainty \citep[e.g.,][]{Leutbecher2017StochasticVision,Zanna2019UncertaintyPredictions,Eidhammer2024An6}. For each of the resultant ocean states, we examine the adjustment of meltwater input at both the surface and at depth in order to capture uncertainty in how the meltwater is prescribed into the ocean.

\begin{figure*}[h!]
  \noindent\includegraphics[width=\textwidth,angle=0]{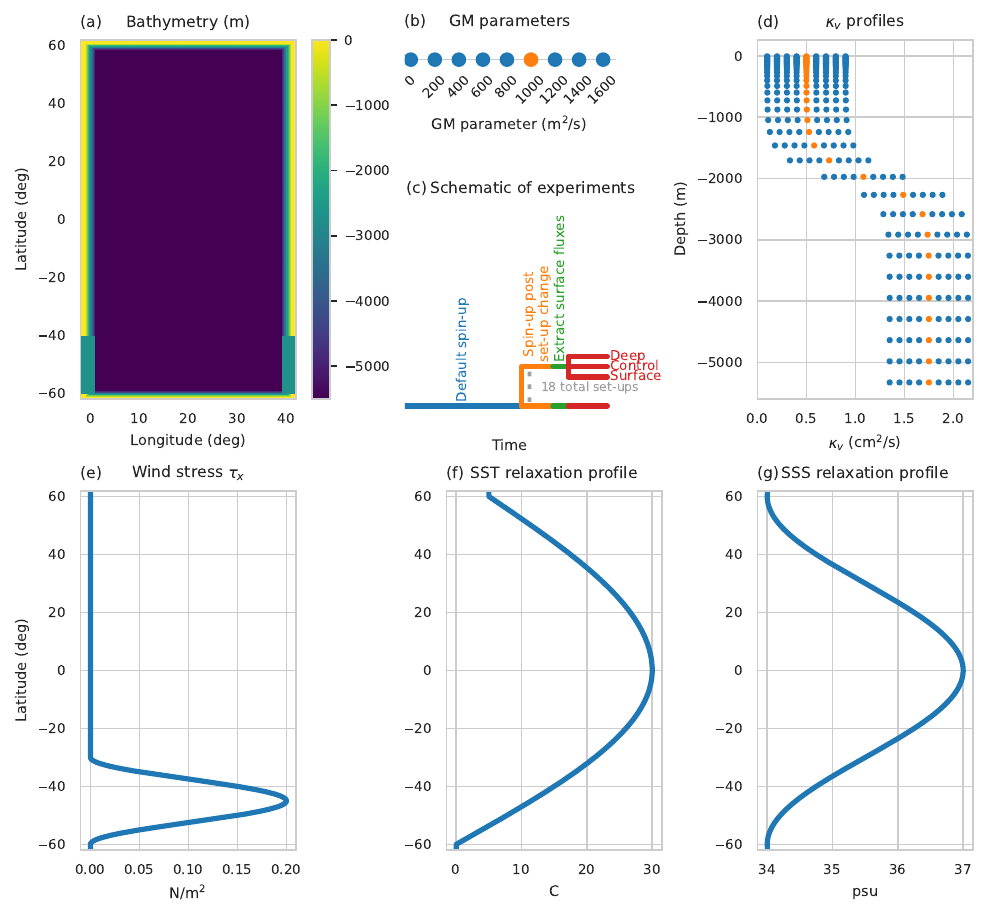}\\
  \caption{Numerical model set-up. (a): The single basin bathymetry, identical to \cite{Eisenman2024TheFluxes}, with linearly sloping continental shelves and a re-entrant channel with a ridge. (b): The values of the $\kappa_{GM/Redi}$ parameter in the simulations. (c) Schematic of experiments as described in text. Note that partway through the spin-up runs (shown in blue and orange), we switched from a linear free surface to a non-linear free surface, because the latter, which is less commonly used, perfectly conserves tracers. For the $\frac{1}{4}^{\circ}$ run, this switch was done at year 3414 (of the total additional 5726 year spin-up), while for the $1^{\circ}$ runs, this was performed at the time of branching. (d) The different vertical diffusivity profiles considered.
  For both the GM parameter and the vertical diffusivity, unless the parameter is specified in text to be changed, the default values highlighted in orange are used in the coarse simulation.
  (e): The zonal wind stress profile, primarily applied in the Southern Ocean region. 
  (f)/(g): The sea surface temperature (SST) and sea surface salinity (SSS) relaxation profiles utilized in the spin-up runs.} \label{fig:model_setup}
\end{figure*}
% Made in fig1_plot.ipynb. Note didn't have to be updated in the fixed runs, but did a clean making of all figures for updated code which is why this is being updated

\section{Methods} \label{sec:methods}

\subsection{Model configuration and spin up} \label{subsec:spinup}

We use the MITgcm ocean model \citep{Marshall1997AComputers} in an idealized single-basin domain which spans from $-62^\circ$ to $62^\circ$ in latitude and $0^\circ$ to $42^\circ$ in longitude. The basin has continental shelves which linearly slope down to the full depth of $5500$m, and includes a re-entrant channel with a ridge at $2750$m depth (see Figure \ref{fig:model_setup}a). This set-up is similar to \cite{Munday2013EddyCurrents}, but with a wider basin and continental shelves. Our single-basin geometry allows for a large number of perturbed parameter runs at lower computational cost than realistic geometry and resembles the domain used in many previous studies \citep[e.g.,][]{Jansen2018TransientWarming,Jones2011TheWinds,Marques2022NeverWorld2:Resolutions}.
\\

We utilize models with both $1^{\circ}$ and $\frac{1}{4}^{\circ}$ horizontal resolution. For both, we use no-slip boundary conditions, an implicit nonlinear free surface, and a convective adjustment scheme which applies a high vertical diffusivity (1 m$^2$/s) to mix unstable density profiles. We employ a nonlinear equation of state which is a modified UNESCO formula by \cite{Jackett1995MinimalStability}. The vertical diffusivity ($\kappa_v$) is a function of depth \citep{Bryan1979AOcean}, with the default profile shown in orange in Figure \ref{fig:model_setup}d. In the $1^{\circ}$ set-up, we use Laplacian horizontal viscosity, whereas in the $\frac{1}{4}^{\circ}$ set-up we use Smagorinsky horizontal biharmonic viscosity. For both resolutions, we use a vertical viscosity coefficient of $3\times 10^{-3}$ m$^2$s$^{-1}$. In the $1^{\circ}$ runs, we use the Gent-McWilliams (GM) parameterization \citep{Gent1990IsopycnalModels} with a constant isopycnal thickness diffusivity set to $1000$ m$^2$s$^{-1}$ by default and the Redi isopycnal tracer diffusivity \citep{Redi1982OceanicRotation} set equal to the GM parameter (Figure \ref{fig:model_setup}b); throughout the text we refer to this as the $\kappa_{GM/Redi}$ parameter. We use no GM or Redi parameterization in the $\frac{1}{4}^{\circ}$ runs. 
\\

We begin with a long spin-up run in $1^{\circ}$ resolution with the default vertical diffusivity profile and $\kappa_{GM/Redi}$ parameter value. The spin-up run imposes zonal wind in the southern part of the domain shown in Figure \ref{fig:model_setup}e and relaxes the sea surface temperature (SST) and sea surface salinity (SSS) to specified profiles (see Figure \ref{fig:model_setup}f/g). The relaxation timescales are set to 10 days and 30 days for SST and SSS, respectively. The $1^{\circ}$ spin-up run is performed for 7540 years until there is equilibration of the deep ocean (see supplementary text S1 in \cite{Eisenman2024TheFluxes}). 
\\

To generate different ocean states, we initialize simulations starting from the default spin-up run, perturb the parameter values or resolution, and re-equilibrate the ocean with the restoring boundary conditions. In total, we create 18 versions of the model including one version at $\frac{1}{4}^{\circ}$ resolution and 17 versions at $1^{\circ}$ resolution with different $\kappa_{GM/Redi}$ parameter values and vertical diffusivity profiles. For the $1^{\circ}$ runs, we branch from the end of the 7540 year run and vary either $\kappa_{GM/Redi}$ (Figure \ref{fig:model_setup}b) or vertical diffusivity (Figure \ref{fig:model_setup}d), while holding the other at the default value. For each change from the default version of the model, we continue a spin-up run until equilibration after introducing the parameter change, with additional spin-up needed for between 200 and 1309 years; smaller parameter changes required less additional spin-up, while larger parameter changes needed more to achieve equilibration. The $\frac{1}{4}^{\circ}$ spin-up run was branched 600 years into the default $1^{\circ}$ spin-up run, with an additional runtime to equilibration of 5726 years after the branching. \\
%can see this from the end of the 3000on.ipynb file where we plot the spinup run. We see that the last of the 6 periods chosen for fluxes ends at year 5701+25. Also the switch from a linear to nonlinear free surface was done 3414 years into this 5726 run.
%you can see how much more spinup was needed at the bottom of the checking_perturbedkv.ipynb and checking_perturbedGM.ipynb files. The range here is the minimum and maximum from which we took fluxes 

Parameter choices sampled in the different versions of the $1^{\circ}$ model are similar to those in \cite{Huber2017DriversUptake} and \cite{Newsom2023BackgroundEfficiency}, who chose parameter values consistent with CMIP5 to sample parametric uncertainty. These parameter choices are consistent with our aim of capturing a wide range of reasonable ocean states through which to understand dependencies of the ocean response to meltwater. We note that the parameters do not aim to fully capture parameter uncertainty across observations or newer state-of-the-art coupled model ensembles. For example, a common assumption, which we follow, is that different Bryan-Lewis profiles of vertical diffusivity have the same functional form and are shifted by a constant at every depth \citep[e.g.,][]{Ehlert2017TheMixing,MacDougall2017TheParameters,Huber2017DriversUptake,Newsom2023BackgroundEfficiency}. This results in our parameter ranges covering less uncertainty in the deep ocean and more in the upper ocean compared to the uncertainty in recent observational estimates \citep[e.g.,][]{Oka2025DeepEstimates}. Similarly, we vary the GM parameter but assume it is spatially constant, while the modeling community tends to favor a spatially varying parameter in some recent studies \citep{Loose2023ComparingModel}. 

\subsection{Surface flux boundary conditions} \label{subsec:fluxbd}
We aim to isolate and focus on the range of ocean dynamic responses to meltwater perturbations, and thus to examine the responses without damping from the surface boundary restoring (the idealized atmosphere). After the spin-up runs, for each model set-up, we prescribe surface fluxes rather than restoring boundary conditions following the protocol design of  \cite{Zika2018ImprovedWarming} and \cite{Todd2020OceanOnlyChange} and as also used in \cite{Eisenman2024TheFluxes}. \\

For each run, we identify a 25-year period from the end of the spin-up run with near $0$ global and temporal mean heat and salt fluxes; in practice, across all 18 model set-ups, the surface fluxes chosen from the end of the spin-up runs would result in volume mean drift of less than $4.0\times 10^{-5}$ K/year and $9.8\times 10^{-6}$ g/kg/year, which are the same orders of magnitude as in \cite{Eisenman2024TheFluxes}. We perform a temporal mean over the identified period to generate a time-constant map of surface heat and (virtual) salt fluxes. This approach is in contrast to previous studies that extracted fluxes from the spin-up run with high temporal frequency (e.g., 6-hourly or daily) and then imposed them as a repeating cycle onto the ocean \citep[e.g.,][]{Todd2020OceanOnlyChange}. In the case of previous literature, coarse resolution models were used, whereas in this work we include a $\frac{1}{4}^{\circ}$ eddy-permitting run, meaning that fluxes at high temporal resolution may be associated with transient eddies in the spin-up run. The temporal averaging approach avoids imposing fluxes with transient eddies from a spin-up run that are not associated with eddies in the flux-forced run. A trade-off of this approach is that imposing a climatology will likely impact the ocean response compared to high frequency fluxes \cite[see][for an analysis of such issues in the case of wind forcing]{Luongo2024RetainingSimulations}. This strategy is repeated for all runs, including the $1^{\circ}$ runs, for consistency. For the eddy-permitting resolution ($\frac{1}{4}^{\circ}$) simulation, we generate an ensemble of 6 members to sample internal variability associated with the presence of eddies, each with its own fluxes found from non-overlapping 25-year periods near the end of the spin-up run. In all flux-forced simulations, we continue to use the same wind profile as in the restoring boundary runs (Figure \ref{fig:model_setup}e).
%The only difference between the approach in the eddy-permitting resolution ($\frac{1}{4}^{\circ}$s) and $1^{\circ}$ resolution is that we independently repeat this process 6 times in the eddy-permitting resolution, such that we can perform all $\frac{1}{4}^{\circ}$ experiments with a 6 member ensemble. 

\subsection{Meltwater injection} \label{subsec:meltwater_injection}
For each parameter set-up, we initiate three runs which all use the constant surface fluxes as outlined in Section \ref{sec:methods}\ref{subsec:fluxbd}. These runs are (a) control with no meltwater input; (b) surface meltwater input; and (c) deep meltwater input. For both perturbation experiments, we add fluid into the ocean uniformly at the southernmost ocean cell (just north of the continental shelf). The surface meltwater experiment uses volume input at all longitudes in the top grid cell in $z$-space (between 0 and 10m depth), while in the deep meltwater experiment, we input water in the grid cell located around 1000m (between 954 and 1137m depth). This approach is similar, though slightly simplified, to studies that represent basal melt in models without a resolved ice-shelf cavity by uniformly injecting freshwater over the vertical extent of the cavity \citep[e.g., the suggested parameterization in ][]{Mathiot2017Explicit3.6}. For both depths of injection, the meltwater input is performed using the ``AddMass" option in MITgcm, which inputs a real, rather than virtual, water flux (i.e., it changes the ocean volume). We input the water at $0^\textrm{o}$C and $0$ psu.
\\

The meltwater input is prescribed as a constant 0.1 Sv spread over all longitudes of the basin. This choice of input is large given the ice sheet mass balance in the current climate, but useful for comparison to previous literature \citep[e.g.,][]{Bronselaer2018ChangeMeltwater, Lago2019ProjectedContributions,Bronselaer2020ImportanceOcean,Beadling2022ImportanceChange,Eisenman2024TheFluxes} and to proposed Tier 1 experiments in \cite{Swart2023TheDesign}.

\subsection{Metrics} \label{subsec:metrics}
In this paper, we focus on a few metrics to quantify the change in ocean states initiated by parameter value and resolution changes, as well as the response of the ocean to meltwater.
\\

We use the meridional overturning streamfunction as a metric for the large scale circulation, $\psi$, which we compute in density coordinates and then remap to $z$-coordinates. In particular, we compute the following integral in potential density coordinates $\sigma$ referenced to the surface (i.e., $\sigma=\sigma_0$):
\begin{equation} \label{eq:densitymoc}%see https://journals.ametsoc.org/view/journals/clim/33/9/jcli-d-19-0215.1.xml and https://agupubs.onlinelibrary.wiley.com/doi/10.1029/2025JC022651
    \psi(y,\sigma,t)=\int^{\sigma}_{\sigma_{\textrm{min}}}\int_{x_w}^{x_e} v(x,y,\sigma',t) dx d\sigma',
\end{equation}
with yearly averaged data and $\sigma_{\textrm{min}}$ equal to the minimum potential density. Here, $x_e$ and $x_w$ are the locations of the eastern and western edges of the basin, respectively. We project $\psi$ back to depth space using the yearly averaged depth (in $z$) of each isopycnal $\sigma$; thus, we remap $\psi(y,\sigma,t)\rightarrow \psi(y,z,t)$. In practice, Equation \eqref{eq:densitymoc} is calculated using the ``layers" package in MITgcm, where $v(x,y,\sigma,t)$ includes both the resolved and the bolus velocity from the GM parametrization (if utilized). The strength of the upper cell meridional overturning circulation (MOC) is computed as the maximum of $\psi$ north of 20$^\circ$N and below 300 meters depth, and the strength of the abyssal cell is the maximum of $|\psi|$ below 2000 meters (at any latitude, and thus typically set in the Southern Ocean). Note that values are sensitive to the coordinate used for streamfunction calculation -- for example, if we had calculated the streamfunction in depth coordinates rather than potential density coordinates, the abyssal cell would be weaker due to cancellation from the Deacon cell \citep{Doos1994TheOcean}.
%Thus, we can find yearly values of the upper and lower cell MOC strengths in each simulation.
\\

% Both the ocean model set-up and meltwater injection affects the stratification of the model, which we calculate using the standard metric of the buoyancy frequency:
% \begin{equation} \label{eq:n2}
%     N^2=-\frac{g}{\rho}\frac{\partial \rho}{\partial z}.
% \end{equation}

% where $\rho$ is density and $g$ is gravity. 

In the meltwater perturbation runs, we examine dynamic sea level which is defined at any point (or area) as the anomaly of the model free surface ($\eta$) from the global mean. We focus on the Northern Hemisphere dynamic sea level as this indicates propagation to the far end of the basin from the meltwater injection location; we note that the Southern Hemisphere dynamic sea level is exactly equal and opposite by definition.
\\

Finally, for some metrics, we quantify the change over perturbed model runs. To do this, we perform a linear fit starting from year 100 of the timeseries, and we compute the value at the final year (year 200) with associated uncertainty utilizing this linear trend and its standard error. The assumption of a linear trend after the first 100 years of adjustment is based on the rationale that although the initial adjustment to perturbations is nonlinear, the longer-term adjustment tends to be approximately linear (e.g., see Figure \ref{fig:timeseries_dynamicsealevel} in the Appendix for timeseries of Northern Hemisphere dynamic sea level in response to meltwater input). The linear fit approach avoids issues due to internal variability that can arise, for example, when taking the mean over the last few decades.

\section{Results} \label{sec:results}
\subsection{Variations in control ocean states} \label{subsec:variationscontrol}

\begin{figure*}[t]
  \noindent\includegraphics[width=38pc,angle=0]{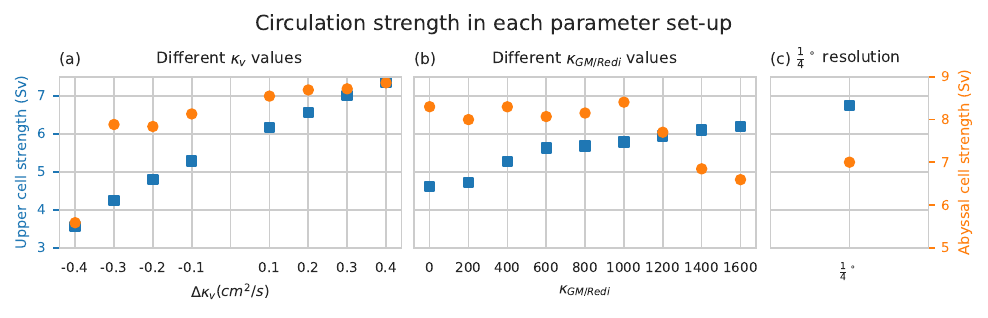}\\
  \caption{The circulation strength, both the upper cell and the abyssal cell, in the control simulations for the perturbed parameter ensemble. (a) for different $\Delta \kappa_v$ values; (b) for different $\kappa_{GM/Redi}$ values; (c) for the $\frac{1}{4}^{\circ}$ resolution.} \label{fig:control_moc_setup}
\end{figure*}

% Made in moc_strength.ipynb. Note didn't have to be updated in the fixed runs, but did a clean making of all figures for updated code which is why this is being updated

\begin{figure*}[t]
  \centering
  \noindent\includegraphics[width=\textwidth,angle=0]
  {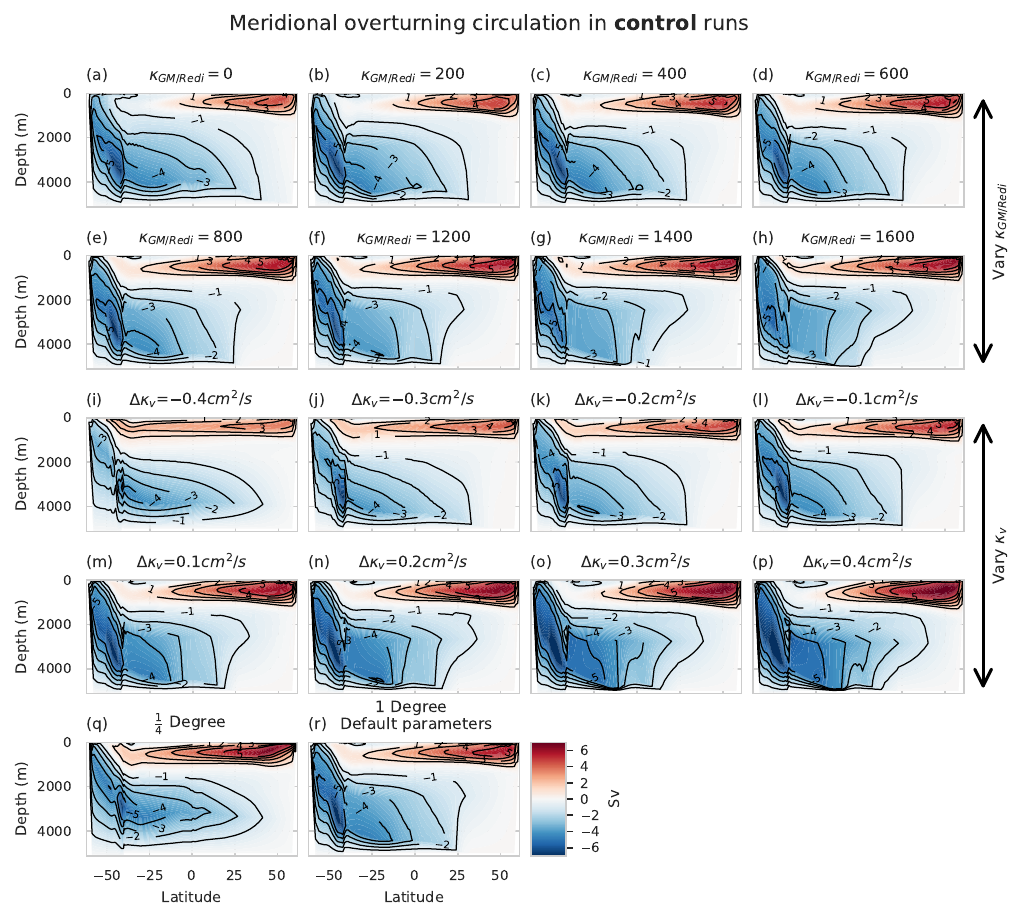}\\
  \caption{Meridional overturning streamfunction in the control runs: (a)-(h) with varying GM/Redi parameter ($\kappa_{\textrm{GM}}, \kappa_{\textrm{Redi}}$) as indicated in Figure \ref{fig:model_setup}b, with the parameter value indicated in each subplot title; (i)-(p) changing the vertical diffusivity as indicated in Figure \ref{fig:model_setup}d, with the value in the subfigure title indicating the shift from the default profile (uniformly over the whole depth); (q): the $\frac{1}{4}^{\circ}$ run; and (r): the 1-degree run with default parameter values (see Figure \ref{fig:model_setup}).} \label{fig:controlstreamfunction}
\end{figure*}

As designed, the parameter and resolution modifications in the ocean model, described in Section \ref{sec:methods}, result in a range of different ocean states. These effects are summarized using the streamfunction strengths in Figure \ref{fig:control_moc_setup}. Here, we find that the $\frac{1}{4}^{\circ}$ simulation has a stronger upper cell MOC and a slightly weaker abyssal cell compared to the default state of the model ($1^{\circ}$ with default parameter values shown in Figure \ref{fig:model_setup}b and d). For the $1^{\circ}$ set-ups, we find that the range of different states due to changing $\kappa_v$ is larger than the range generated due to changing the $\kappa_{GM/Redi}$ values, including the impact on both the upper and abyssal cell strength. Increasing $\kappa_v$ strengthens both the upper \citep[as also seen in][]{Huber2017DriversUptake} and abyssal cells (see also Figure \ref{fig:controlstreamfunction}). The strengthening of the abyssal cell with $\kappa_v$ is consistent with \cite{Nikurashin2011AOcean} and \cite{Stewart2014OnOcean} who argue that the cell strengthens with increased $\kappa_v$ due to an advective-diffusive balance. Increasing $\kappa_{GM/Redi}$ results in a strengthened upper cell and small variations in the abyssal cell strength. This result for the upper cell is counter to \cite{Marshall2017TheStudy,Saenko2018ImpactModel,Huber2017DriversUptake}, which in global model set-ups have found that increasing the GM parameter \textit{decreases} the MOC strength. Given our simplified model set-up and forcing profiles, it is likely that there are different controls on the strength of the overturning circulation here compared to realistic geometry. In particular, we see that the bolus streamfunction, corresponding to the contribution from the GM parameterization, follows the expected response in the Southern Ocean \citep[e.g.,][]{Saenko2018ImpactModel} -- namely, the strength of the bolus streamfunction increases with increasing $\kappa_{GM/Redi}$ (see Figure S1 in the Supplementary Materials). However, the increase in the residual MOC strength with larger $\kappa_{GM/Redi}$ appears to be driven by a stronger contribution at the deep water formation site at the northern edge of the basin, which may be linked to simplified boundary conditions and a lack of any imposed wind driven Ekman transport in this location. Despite this, we continue to use the simulations ranging $\kappa_{GM/Redi}$ as they behave as expected in the Southern Ocean, which is key for this study due to the location of meltwater input. Additionally, these simulations contribute to the broad goal of generating a range of background states to identify physical dependencies.
\\

% Made in streamfunctions_and_n2.ipynb. Note didn't have to be updated in the fixed runs, but did a clean making of all figures for updated code which is why this is being updated

The ocean stratification changes such that larger $\kappa_v$ typically leads to a decrease in stratification in the upper ocean (top $\sim$100 meters in the midlatitudes) and an increase in stratification below that. In the Southern Ocean, the decrease in upper ocean stratification  (and vice versa for smaller vertical diffusivities) extends deeper, down to $\sim$800 meters. These general trends have complex structures with additional changes at depth that have latitudinal dependence (Figure \ref{fig:control_n2} in the Appendix). Changes to the $\kappa_{GM/Redi}$ parameter values result in smaller changes in the stratification compared to changing $\kappa_v$, although low values of the parameters tend to lead to an increase in the stratification in the upper Southern Ocean (e.g., Figure \ref{fig:control_n2}a-c). In the $\frac{1}{4}^{\circ}$ simulation, we find a weaker stratification near the surface and stronger stratification below that in the midlatitudes compared to the default $1^{\circ}$ model; in the Southern Ocean, we find stratification changes in the upper ocean which are dependent on the latitude and have a complex vertical structure.
\\

Overall, we find that these simulations sample a relatively wide spread of distinct possible states by changing the ocean model parameter values and resolution. This is indicated by the spread in the upper cell MOC, for example, where the strength in the default $1^{\circ}$ model was 5.78 Sv and the sampled range has limits of 3.56 and 7.36 Sv; thus, the maximum spread from the default parameter values is 38\% (37\% compared to the ensemble mean). This is comparable to the AMOC range sampled in \cite{Huber2017DriversUptake} due to differing parameter values of 34\%. It is also comparable to the spread in CMIP6, which is 34\% compared to the ensemble mean \citep[calculated from Figure 1 of][]{Nayak2024ControlsModels}. The perturbed parameter ensemble also samples spread in the abyssal MOC and the stratification. %Finally, we note that, as expected, the $\frac{1}{4}^{\circ}$ simulation does not project onto any of the $1^{\circ}$ runs and we have sampled a variety of distinct ocean states.
\\

\subsection{Ocean Circulation Responses} \label{subsec:transport}

\begin{figure*}[t]
  \centering
  \noindent\includegraphics[width=\textwidth,angle=0]{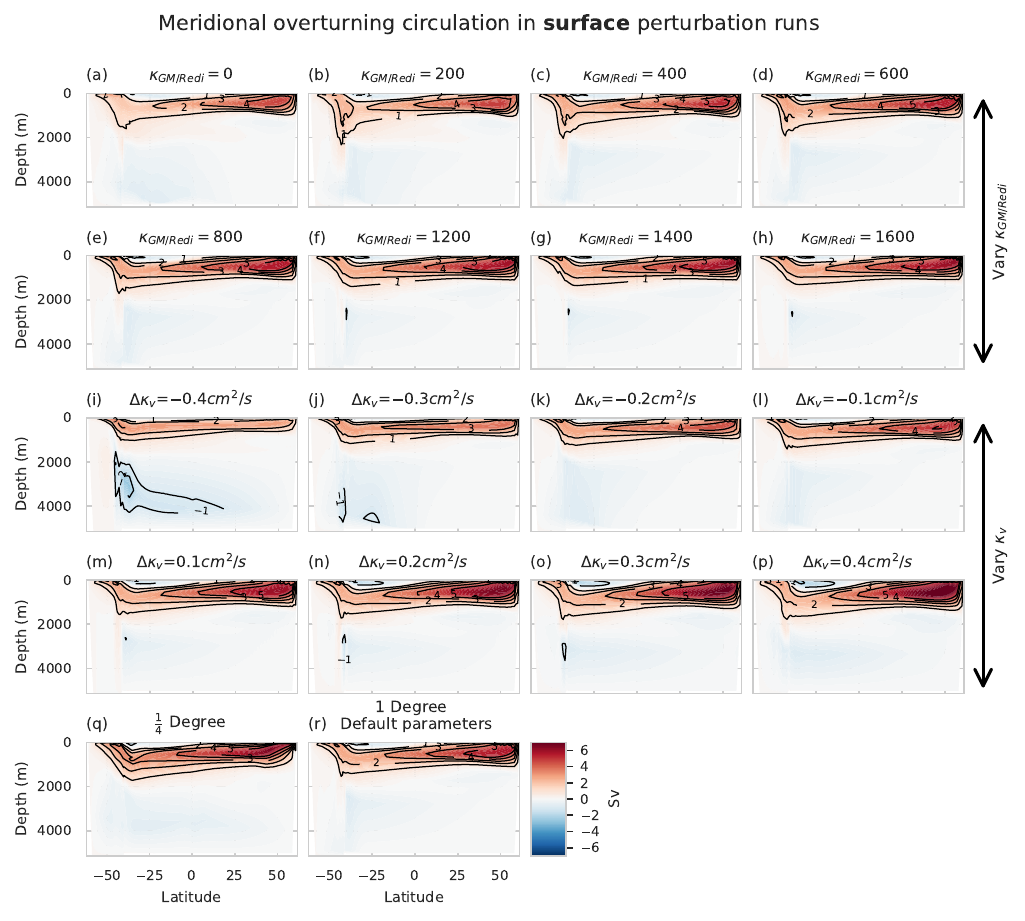}\\
  \caption{Meridional overturning streamfunction in the surface perturbation runs: (a)-(h) with varying GM/Redi parameter ($\kappa_{\textrm{GM}}, \kappa_{\textrm{Redi}}$) as indicated in Figure \ref{fig:model_setup}b, with the parameter value indicated in each subplot title; (i)-(p) changing the vertical diffusivity as indicated in Figure \ref{fig:model_setup}d, with the value in the subfigure title indicating the shift from the default profile (uniformly over the whole depth); (q): the $\frac{1}{4}^{\circ}$ run; and (r): the 1-degree run with default parameter values (see Figure \ref{fig:model_setup}).} \label{fig:surfacestreamfunction}
\end{figure*}
% Made in streamfunctions_and_n2.ipynb. 

\begin{figure*}[t]
    \centering
  \noindent\includegraphics[width=\textwidth,angle=0]{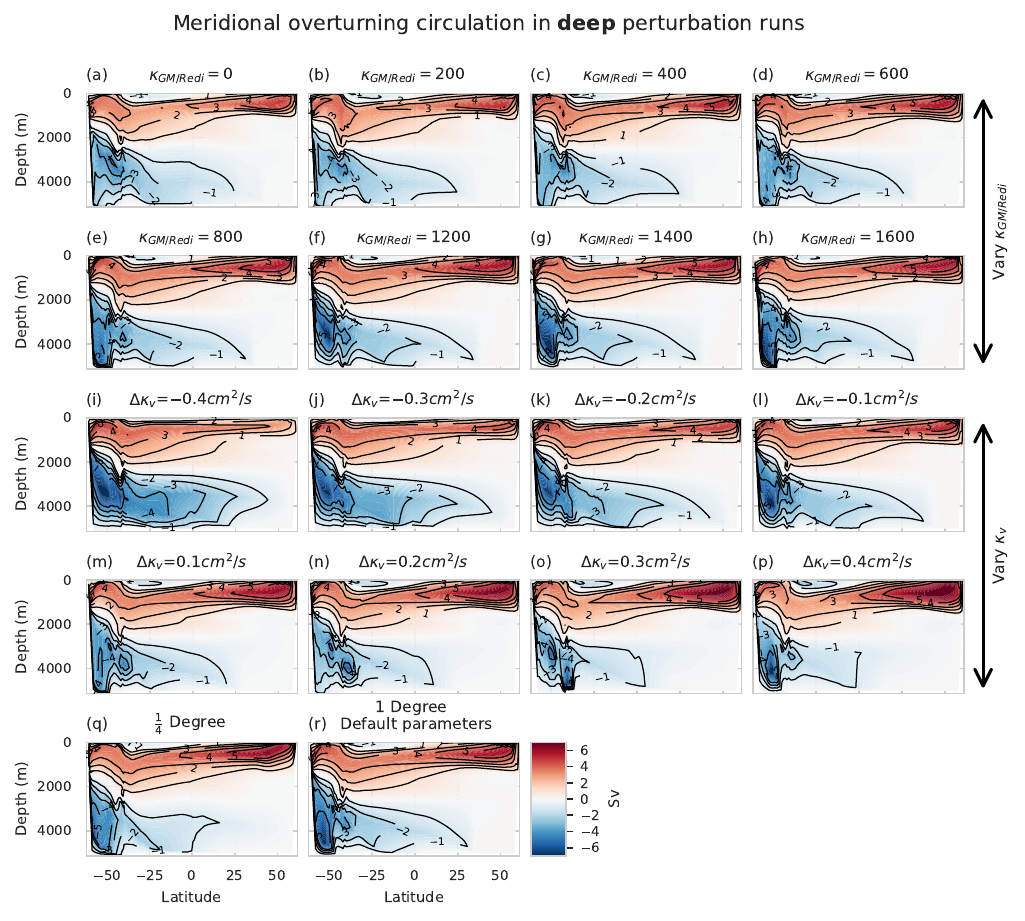}\\
  \caption{Meridional overturning streamfunction in the deep perturbation runs: (a)-(h) with varying GM/Redi parameter ($\kappa_{\textrm{GM}}, \kappa_{\textrm{Redi}}$) as indicated in Figure \ref{fig:model_setup}b, with the parameter value used indicated in each subplot title; (i)-(p) changing the vertical diffusivity as indicated in Figure \ref{fig:model_setup}d, with the value in the subfigure title indicating the shift from the default profile (uniformly over the whole depth); (q): the $\frac{1}{4}^{\circ}$ run; and (r): the 1-degree run with default parameter values (see Figure \ref{fig:model_setup}).} \label{fig:deepstreamfunction}
\end{figure*}
% Made in streamfunctions_and_n2.ipynb. 
%The overturning streamfunctions, computed in density coordinates, after the introduction of meltwater are shown in Figures \ref{fig:surfacestreamfunction} and \ref{fig:deepstreamfunction} for the surface and deep input experiments,  respectively. We additionally focus on timeseries of the abyssal MOC response in Figure \ref{fig:abyssal_cell_trend}. Timeseries of the upper cell MOC are shown in the appendix in Figure \ref{fig:upper_cell_trend}, although we focus primarily on the change in spatial structure (Figure \ref{fig:surfacestreamfunction}) as the upper cell strength metric is primarily set in the North Atlantic, rather than the Southern Ocean.
%\\

The most robust response to the surface meltwater forcing is a near total shutdown of the abyssal cell, due to stabilized stratification near the ocean surface which limits deep water formation; this behavior has been noted in previous studies of the response to Antarctic meltwater input \citep[e.g.,][]{Lago2019ProjectedContributions,Li2023AbyssalMeltwater,Moon2025AntarcticCirculation}. The timeseries of the abyssal MOC strength for surface input experiments (Figure \ref{fig:abyssal_cell_trend}a and c) show consistent abyssal MOC strength reduction over the first 100-150 years with no recovery after this period of weakening. This holds for all set-ups, but there is quantitative dependence on the $\kappa_v$ and $\kappa_{GM/Redi}$ values used. In particular, runs with a weaker abyssal cell in the control run tend to weaken less. For example, the lowest vertical diffusivity run, which also has the weakest control abyssal circulation, weakens much less than the other runs (see also Figure \ref{fig:surfacestreamfunction}). 
\\

Meltwater forcing at depth also induces substantial changes in the abyssal cell. However, rather than a near shutdown of the circulation, as in the case of surface meltwater input, the abyssal cell is maintained. As in the control simulation, the abyssal cell is closed by the formation of deep water, which in our simulations occurs at the southern-most latitudes with approximate zonal uniformity. However, the outcropping region is much smaller than in the control simulation and is not visible in the streamfunction as plotted, because it narrowly occurs along the continental shelf. 
%North of 60S, in the last 50 years of the deep meltwater input runs, the shallowest that the abyssal cell reaches is $\sim$ 1000-1500 meters in the Southern Ocean (Figure \ref{fig:deepstreamfunction}). 
\\

\begin{figure}[t]
    \centering
  \noindent\includegraphics[width=0.5\textwidth,angle=0]{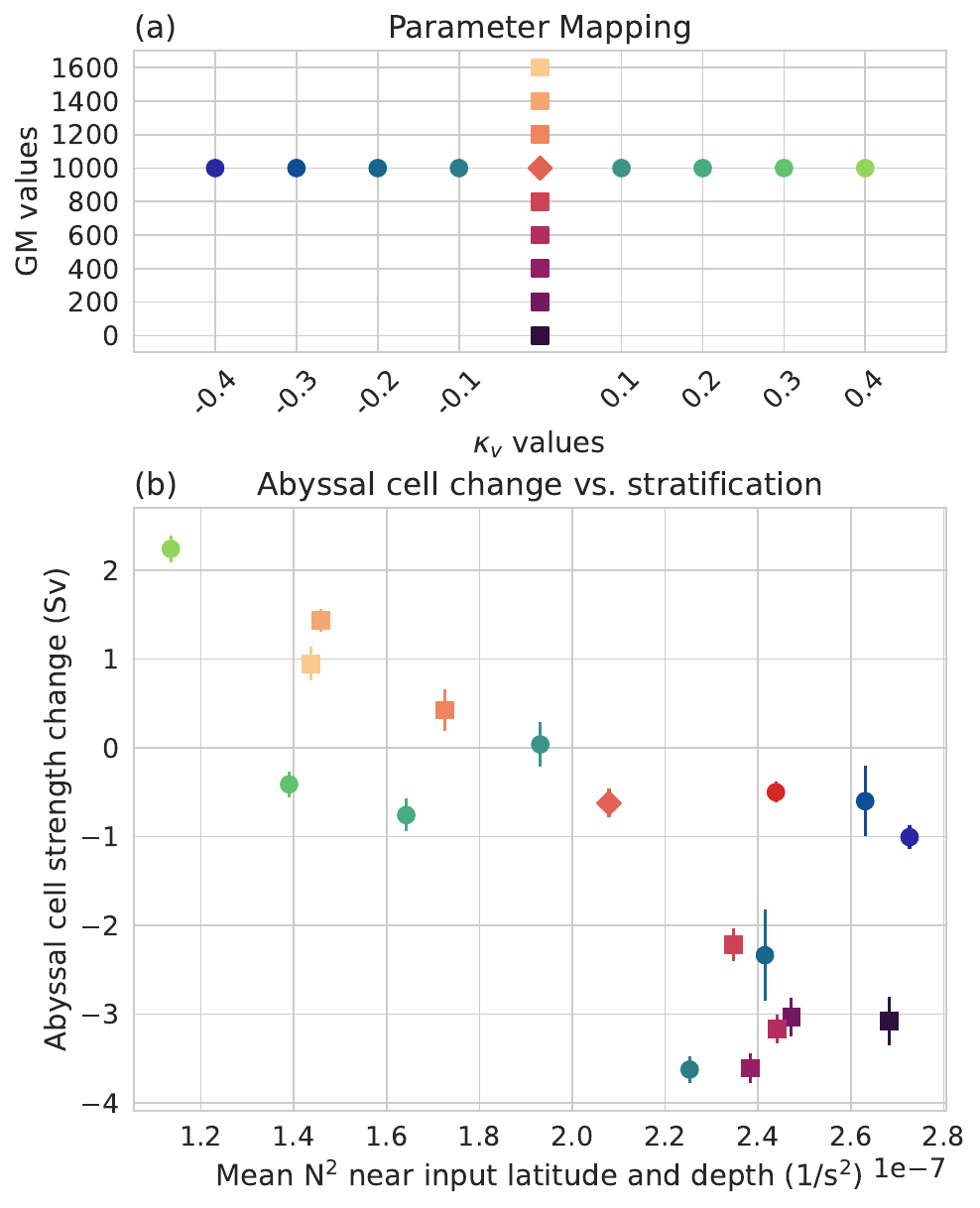}\\
  \caption{Change in the abyssal cell strength in the deep perturbation runs compared to the stratification near the injection depth and location in the control simulation (panel b). The color coding is shown in panel (a), except for the $\frac{1}{4}^{\circ}$ simulation which is shown in red. We evaluate the mean from 650 meters depth to 1150 meters depth (based on the grid spacing in the vertical) and from the continental shelf to 55S. } \label{fig:deepstreamfunction_correlation}
\end{figure}

\begin{figure*}[t]
  \noindent\includegraphics[width=\textwidth,angle=0]{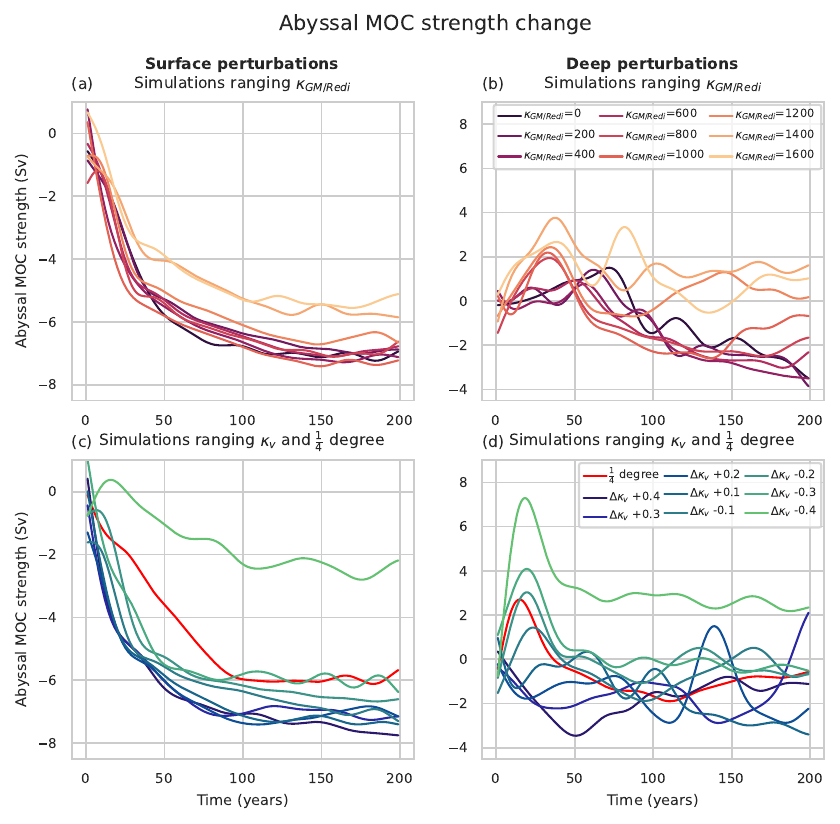}\\
  \caption{The strength of the abyssal cell in response to meltwater perturbations compared to the control. The legend for panels (a) and (b) is shown on panel (b), while the legend for panels (c) and (d) is shown on panel (d). Panels (a) and (c) at the surface; (b) and (d) at depth. Curves are smoothed with a 6th order low-pass Butterworth filter with a cut-off period of 30 years. The top row (a and b) are for different $\kappa_{GM/Redi}$ parameters and the bottom row (c and d) are for varying $\kappa_v$. The $\frac{1}{4}^{\circ}$ results are shown in (c) and (d).  Here, we consider the MOC \textit{change}, obtained by subtracting the mean of the appropriate control run. As in Figure \ref{fig:control_moc_setup}, the abyssal MOC strength is found for the absolute value (so negative here in the change plot means a weakening).} \label{fig:abyssal_cell_trend}
\end{figure*}

In the deep perturbation runs, the dynamics closing the abyssal cell differ from the control simulations. In the control simulations, the scaling of the abyssal cell follows theories such as \cite{Nikurashin2011AOcean}, where the steady-state residual overturning is understood through a match of approximate adiabatic dynamics in the Southern Ocean channel and diabatic dynamics in the basin, resulting in an advective-diffusive balance.
The streamfunction in the control simulations scales monotonically with $\kappa_v$, which is consistent with \cite{Stewart2014OnOcean}. However, in the case of the adjustment to deep meltwater input, the dynamics vary from the control case. The introduction of meltwater at depth causes substantial mixing through convective adjustment, and the dynamics in the channel are transient and diabatic as the plume rises, entraining water around it and mixing up to a neutral buoyancy. The strength of the abyssal MOC after meltwater introduction varies between ocean model set-ups (Figure \ref{fig:abyssal_cell_trend}b and d). This abyssal cell change may be related to how effectively the plume entrains water up towards the surface to interact with surface fluxes and produce deep water. This is expected to be related to the stratification near the input location, as mixing only occurs for a convectively unstable profile; see Figure \ref{fig:deepstreamfunction_correlation}, which compares the abyssal cell strength change with the control simulation stratification near the injection location ($R=-0.75$). 
\\

% Made in deep_cell_strength.ipynb. 

The changes in the upper cell in the meltwater injection simulations are smaller than the changes in the abyssal cell. In each surface perturbation experiment, the upper cell is stronger in the Southern Ocean compared to the relevant control experiment, with more pronounced increases for low $\kappa_{GM/Redi}$ values. The extended upper cell visible in the streamfunctions is a result of the abyssal cell (negative circulation) not outcropping in the Southern Ocean and may also be linked to increased export of water associated with meltwater input. Across model set-ups, there are differences in the strength and depth of the upper cell north of the idealized Southern Ocean, which are correlated with the upper cell in the control state (Figure \ref{fig:surfacestreamfunction}); weaker control state MOCs for different $\kappa_v$ values tend to lead to more weakening under forcing, but the relative change compared to the control overturning is small for all parameter sets (less than 15\% change). In the deep perturbation experiment, the upper cell extends further south than in the control simulation, and it is present at all latitudes north of the plume closing the abyssal cell (Figure \ref{fig:deepstreamfunction}). However, there is a substantially different structure than in the surface perturbation experiment (compare to Figure \ref{fig:surfacestreamfunction}), with a positive streamfunction in the Southern Ocean indicating that meltwater input at depth rises close to the surface between approximately 60S and 50S. Outside of the Southern Ocean, in the Southern Hemisphere midlatitudes, the upper cell is much deeper than in the accompanying surface perturbation experiments.

% Made in moc_strength

\subsection{Dynamic sea level responses to meltwater} \label{subsec:sealevel}
\begin{figure*}[t]
  \noindent\includegraphics[width=\textwidth,angle=0]{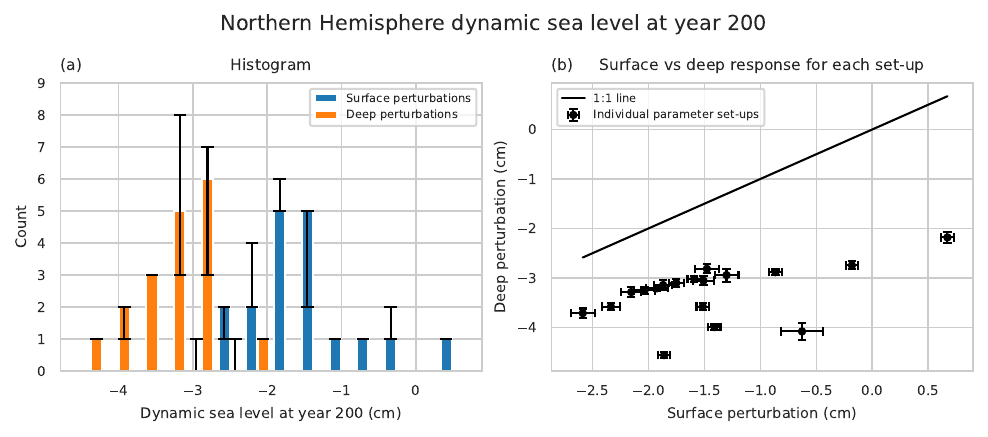}\\
  \caption{Northern Hemisphere mean dynamic sea level at year 200. By definition, an equivalent plot for the Southern Hemisphere would be the exact negative of these results. All values at year 200 are estimated, with associated uncertainty (denoted with the black bars), using a linear trend fit to the timeseries of Northern Hemisphere dynamic sea level between years 100 and 200; this period corresponds to approximately linear behavior, as described in Section \ref{sec:methods}\ref{subsec:metrics}. (a): A histogram of Northern Hemisphere dynamic sea level (with error) for each experiment. (b): Scatter plot of the Northern Hemisphere dynamic sea level in the surface perturbation experiment versus the deep perturbation experiment for each model parameter set-up; all individual points fall below the 1:1 line, indicating that for a given model set-up, the response at the opposite end of the basin from the injection location is always larger in a surface perturbation experiment than in a deep perturbation experiment.} \label{fig:histogram}
\end{figure*}

The response of the dynamic sea level in the Northern Hemisphere to meltwater perturbations for each model set-up is shown in Figure \ref{fig:histogram}a. We find that perturbations at depth result in smaller responses at the far end of the basin from the input compared to surface perturbations, consistent with \cite{Eisenman2024TheFluxes}. In particular, for any individual ocean model parameter set-up, we find that the Northern Hemisphere dynamic sea level response (after 200 years) to a surface perturbation is more positive than the response to a deep perturbation, indicated by all points lying below the identity line in Figure \ref{fig:histogram}b.

Dynamic sea level can be separated into a mass change component and a steric component using
\begin{equation}
    \zeta' = \underbrace{\frac{p_b'}{\rho_0 g}}_{\text{mass change}} \underbrace{-\frac{1}{\rho_0}\int_{-H}^\zeta \rho' dz}_{\text{steric}},
\end{equation}
where $\rho$ is density, $\rho_0$ is a reference density, $g$ is gravity, $p_b$ is the pressure at the ocean floor (found through hydrostatic balance), and $H$ is the full ocean depth. Here, primed quantities represent the anomaly from a control simulation, with the global mean subtracted. The dynamic sea level pattern is nearly entirely from the slower steric component, whereas the barotropic response adjusts the sea level around the basin rapidly and mostly uniformly. The larger (less negative) dynamic sea level responses in the Northern Hemisphere to surface perturbations, leading to more uniform sea level between the two hemispheres, is linked specifically to the upper ocean steric contribution (see Figure S2 in the Supplementary Materials).
\\

We find spread in the resultant dynamic sea level linked to the parameter set-up of the ocean model alone. This spread leads to overlap in the distributions of sea level change associated with surface versus deep meltwater input, such that Northern Hemisphere dynamic sea level changes of approximately –2 to –3cm at year 200 could result from either forcing depth, depending on the model configuration. The ensemble-mean dynamic sea level response in the Northern Hemisphere (or equivalently, the negative of the Southern Hemisphere change) is $-1.47\pm0.73$ cm for surface perturbations and $-3.27\pm0.52$ cm for deep perturbations.  We note that we used the Northern Hemisphere dynamic sea level at year 200 as a summary metric for the spatial pattern of sea level rise. This quantity is relatively small when compared to the global mean sea level rise at year 200 (11.83m in this domain). However, dynamic sea level changes are more localized and transient, therefore leading to smaller signals when averaged (see Figure \ref{fig:dynamicsealevelpattern} for the spatial pattern of dynamic sea level change). For example at year 50, the global mean sea level rise is 2.91m, but individual locations have dynamic sea level differences of up to 1.03m in the standard 1 degree set-up. %Similarly, individual locations have differences between the deep and surface perturbation experiments of up to 0.24m. All calculated in more_sealevel.ipynb in analysis_2026_revisions
\\

%made in sealevel_plots

\begin{figure}[t]
  \noindent\includegraphics[width=0.5\textwidth,angle=0]{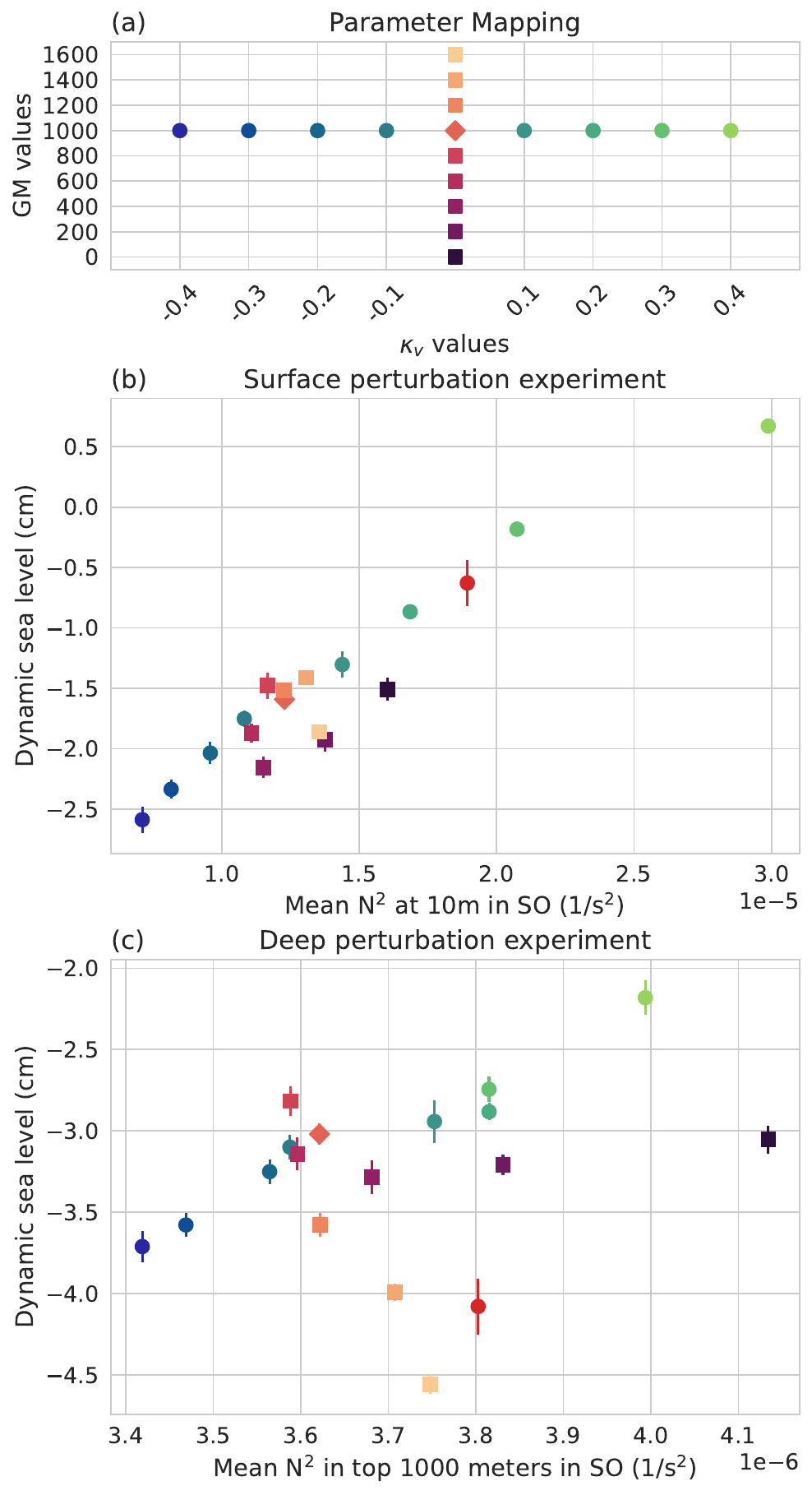}\\
  \caption{Relationship between the Southern Ocean (SO) stratification ($N^2$), defined between 62S (the continental shelf) and 45S and the Northern Hemisphere dynamic sea level response at year 200. The color coding is as defined in panel (a), except for the $\frac{1}{4}^{\circ}$ simulation which is shown in red. (b): Relationship between the dynamic sea level response to a surface meltwater perturbation and the near-surface (10 m depth) stratification in the control run in the SO. %Here the SO is defined by -55 S to -45S, and near-surface stratification focuses on $N^2$ at 10m depth. 
  The correlation is $R=0.96$. (c): Relationship between the dynamic sea level response to a deep meltwater perturbation and the mean stratification in the top 1000m in the control run in the SO. The correlation is $R=0.30$. Note that an equivalent plot for the Southern Hemisphere would be the same except with the negative of the values on the y-axis.} \label{fig:strat_vs_response}
\end{figure}

%\textbf{This is the old subsection d which i am trying to incorporate earlier on}
As discussed in Section \ref{sec:results}\ref{subsec:variationscontrol}, the parameter and resolution changes substantially altered the background state of the ocean. In our experiments with surface meltwater input, there is a positive relationship ($R=0.96$) between the near-surface stratification (at 10m depth) in the Southern Ocean and the dynamic sea level response at the opposite end of the basin (Figure \ref{fig:strat_vs_response}a). We hypothesize that simulations with stronger near-surface Southern Ocean stratification, in the region where sloping isopycnals leads to subduction into the pycnocline, results in injected meltwater subducted to a shallower depth in the low latitudes (and vice versa with weaker stratification). This hypothesis is visible in Figure \ref{fig:density_surface}, which shows the difference in zonally averaged density between the surface perturbation experiment and the control experiment. In experiments with faster adjustment (i.e., lower vertical diffusivities, as in panels i and j), the lower density waters associated with the meltwater stay closer to the surface both in the Southern Ocean and as they enter the low latitudes, while in experiments with slower adjustment (higher vertical diffusivities, panels o and p), the lower density water is more spread out and at lower depths. The resultant faster adjustment of model set-ups where the perturbation remains close to the surface is consistent with theoretical arguments proposed in \cite{Basinski-Ferris2025AEvolution} using a reduced gravity model, constructed to represent dynamics outside the Southern Ocean, which highlighted that faster baroclinic Rossby waves near the surface can help explain faster adjustment of the upper ocean with shallower volume perturbations compared to deeper perturbations. 
\\
%This correlation uses the stratification in the midlatitude Southern Ocean region where sloping isopycnals lead to subduction of water into the pycnocline. Thus, we hypothesize that simulations with stronger near-surface Southern Ocean stratification have the injected meltwater subducted to a shallower depth in the low latitudes (and vice versa with weaker stratification). 

In the case of meltwater injected at depth, the meltwater input is statically unstable and triggers convection, ultimately being spread throughout the top 1000 meters between 60S and 50S (see Figure \ref{fig:deepstreamfunction}, discussed more in Section 3c). Thus, examining the stratification just near the input depth as we did in the surface perturbation experiment is not physically well motivated, as it is unclear what depth of stratification is relevant for determining the depth at which meltwater gets subducted into the pycnocline. We examine the simplest metric of the mean stratification over the upper ocean (top 1000 meters) and find a weak positive correlation ($R=0.30$) between the dynamic sea level response and the mean buoyancy frequency (Figure \ref{fig:strat_vs_response}b), implying that there is not as clear of a connection between the background stratification and the spread of responses as in the surface perturbation experiment.
\\

While we have focused on discussing correlations with stratification and proposed mechanisms that could lead to that correlation, a more thorough mechanistic proof-of-processes is necessary. In particular, cross-correlations can be present between different metrics of the ocean state such as the AMOC strength, the Southern Ocean stratification, or the dynamics of eddies; these cross-correlations can lead to different drivers identified for quantities of frequent investigation \citep[e.g., ocean heat uptake efficiency;][]{Winton2014HasSensitivity,Saenko2018ImpactModel,Liu2023TheSalinity,Newsom2023BackgroundEfficiency}. In our case, for example, the Northern Hemisphere sea level is additionally anti-correlated with the control state AMOC strength for both depths of perturbation (R values of -0.67 and -0.70 for a surface and deep perturbation respectively), despite different directions of background AMOC flow at each depth (Figure S3 in the Supplementary Materials). Thus, other correlations are present in our model simulations, which also highlight the dependence of ocean adjustment on the background state, but need to be separated to make causal arguments.

% made in sealevel_plots

%Thus, we find that the adjustment of the dynamic sea level in response to the meltwater input is dependent on both the ocean model set-up and the location  of the forcing.  In Section \ref{sec:results}\ref{subsec:explanation_stratification}, we relate the variance due to the different ocean model set-ups to the Southern Ocean state in the respective control simulations.

\section{Discussion} \label{sec:discussion}
As mass loss from ice sheets, including Antarctica, is projected to continue over the next century \citep[e.g,][]{Naughten2023UnavoidableCentury}, understanding and correctly representing the ocean dynamic response is important for projections of a range of key climate variables. This includes the regional response of sea level \citep[e.g.,][]{Stammer2008ResponseMelting,Lorbacher2012RapidMelting,Kopp2010TheExperiments,Schmidt2023AnomalousForcing}, as well as the response of large-scale ocean circulation systems \citep[e.g., ][]{Lago2019ProjectedContributions,Li2023AbyssalMeltwater,Moon2025AntarcticCirculation}, and potentially the coupled climate response including impacts on global mean temperature change \citep[e.g.,][]{Dong2022AntarcticEffect, Bronselaer2018ChangeMeltwater,Armour2024Sea-surfaceSensitivity}. 
\\

In this study, we have presented an investigation of potential controls on differing ocean responses to meltwater input. We have used an idealized set-up to create a perturbed parameter ensemble with 18 different configurations, including a resolution change, to partially address parameteric and structural model uncertainty. Given that the model representation of ocean circulation is sensitive to grid resolution and physical parameter values, this ensemble allows us to identify controls on the range of ocean responses to meltwater input that may be linked to the ocean background state alone. We have additionally included idealized meltwater input at two depths due to uncertainty in how meltwater is distributed and motivated by previous work demonstrating a strong sensitivity of the ocean response to the depth of meltwater injection \citep{Eisenman2024TheFluxes,Basinski-Ferris2025AEvolution}. Taken together, we investigate potential important controls on the range of ocean responses to Antarctic meltwater, including uncertainty that may be introduced in projections from both the model ocean state and from interfacing the ocean model with ice sheet fluxes. 
\\

Part of our investigation focused on injection of meltwater at depth, given that a substantial proportion of Antarctic melt is from basal melt under ice shelves, and most current climate models do not resolve ice shelf cavities nor ice-ocean interactions. To approximate the possible uncertainty associated with melt being concentrated at depth from basal melt rather than at the surface from iceberg melt, in our deep meltwater experiments we injected freshwater at around 1000m depth. While simplified, this is similar to the proposed parameterization of \cite{Mathiot2017Explicit3.6}, which suggests injecting freshwater uniformly over the depth of the unresolved ice-shelf cavity. This generates a plume along the vertical wall of the continental shelf, which differs from the plume that would exit along the base of the ice shelf if an ice-shelf cavity was present. This plume drives a buoyant overturning circulation that entrains warm and salty circumpolar deep water, which eventually mixes into the colder and fresher surface layers outside the cavity \citep{Mathiot2017Explicit3.6}. Ultimately, the parameterization produced a near-shelf circulation that was similar to a comparable simulation resolving the ice shelf cavity and entrained the same water masses \citep{Mathiot2017Explicit3.6}. In the present work, for a range of ocean model parameter values, we have shown that the ocean's response to meltwater forcing is highly dependent on whether meltwater is input at depth or at the surface, as shown for a single parameter regime in \cite{Eisenman2024TheFluxes}. In the case of the sea level adjustment, the two depths of meltwater perturbation largely resulted in distinct distributions of Northern Hemisphere dynamic sea level. In particular, for any given ocean model set-up, the surface meltwater input adjusted more quickly around the basin than deep meltwater input. Similarly, the abyssal cell, which changes substantially in response to meltwater injection, is strongly dependent on the depth of the meltwater input. In the case of surface meltwater injection, the cell nearly shuts down due to stabilized stratification near the surface preventing bottom water formation \citep[similar to previous work, e.g., ][]{Lago2019ProjectedContributions,Li2023AbyssalMeltwater,Moon2025AntarcticCirculation}. Unlike in the surface meltwater case, the near-surface stratification is not strengthened under deep meltwater injection, which allows the abyssal cell to close.
\\
%\cite{Pauling2017Time-DependentModel} demonstrated that the impact of latent heat extraction had an effect on the sea ice response in particular, and thus, this possible sensitivity should be investigated in a coupled framework.

The ocean response was demonstrated to depend on physical parameter values such as vertical and eddy diffusivities, as well as model resolution, for both depths of meltwater injection. For example, the adjustment of dynamic sea level across the basin has spread depending on the configuration of the model. In the case of the surface input simulation, this is strongly correlated with the background stratification of the Southern Ocean, which is consistent with the hypothesis that this stratification is related to the subduction of meltwater into the pycnocline in low latitudes. In particular, stronger near-surface stratification results in meltwater remaining closer to the surface and faster adjustment, potentially related to the baroclinic Rossby wave speed, as highlighted in \cite{Basinski-Ferris2025AEvolution}. Similarly, in simulations where the abyssal cell is maintained (i.e., meltwater input at depth), the change in the cell is strongly dependent on the ocean model parameter set-up. We hypothesize that this may be linked to how effectively the plume entrains water and moves upwards, which is expected to depend on the background density profile. We demonstrate that more strongly stratified water columns (in the control state) near the injection location are correlated to a weakening of the abyssal cell while weakly stratified columns are correlated to a strengthening of the abyssal cell. In both cases, we stress that given the cross-correlated metrics of the ocean state, other metrics may also be correlated with the resultant sea level or abyssal cell change, necessitating additional studies aimed at determining causality rather than correlation. 
\\

We have presented evidence that there are drivers of spread of the ocean response to meltwater depending on how the ocean model is configured and the background state that is captured, which may translate to sensitivities in more realistic models. In particular, this work demonstrates that even when controlling for the same freshwater fluxes (which can be prescribed in varying ways, as discussed in \cite{Swart2023TheDesign}) and removing atmospheric feedbacks, the choice of ocean model influences the responses, as has been shown previously for surface fluxes \citep{Huber2017DriversUptake,Todd2020OceanOnlyChange}. This suggests underlying sources of uncertainty that may persist in more realistic models, even if we cannot directly compare our identified controls (e.g., stratification) to similar values in CMIP models due to our idealized geometry. 
\\

Our results linking the spread in meltwater propagation to the ocean model's background Southern Ocean stratification adds to broader literature which has demonstrated that forced responses can be understood or constrained based on the background state, including for ocean heat uptake \citep[e.g.,][]{Newsom2023BackgroundEfficiency,Liu2023TheSalinity,Bourgeois2022Stratification55S} and changes in the Atlantic MOC \citep[e.g.,][]{Gregory2005AConcentration,Bonan2025ObservationalWeakening}. These relationships between the background state and the forced response have previously been used as emergent constraints \citep[e.g.,][]{Bourgeois2022Stratification55S,Liu2023TheSalinity,Bonan2025ObservationalWeakening}, which could be one possible avenue for investigation if the relationship between the Southern Ocean background stratification and the response to meltwater identified here holds in more realistic models. 
\\

While this work focused on the ocean-only response to meltwater, it should be considered in the context of previous studies of the coupled climate response to meltwater input, including interactions with the atmosphere and cryosphere. Antarctic meltwater has been demonstrated to induce a robust response in many components of the climate system, including affecting sea ice extent and surface temperature \citep[e.g.,][]{Ma2011GlobalOcean,Bronselaer2018ChangeMeltwater,Dong2022AntarcticEffect,Li2023GlobalStudy,Sadai2020FutureWarming}, which feedback onto the ocean state. These temperature and sea ice changes are linked to the changed Southern Ocean stratification and heat uptake  \citep[e.g.,][]{Bronselaer2018ChangeMeltwater,Dong2022AntarcticEffect,Li2023GlobalStudy}, among other factors. Thus, the dependence of stratification changes on meltwater injection depth and ocean model set-up, as found here in ocean-only simulations, may impact these results. On the other hand, the dynamic response captured in our ocean-only framework is not expected to be the same as in a coupled atmosphere-ocean-sea ice system where anomalous feedbacks will modify surface fluxes \citep[e.g.,][]{An2024AntarcticTeleconnections}. These coupled feedbacks will alter the ocean-only results and could either amplify or dampen differences between the responses in different ocean model set-ups that we identified. Thus, this study raises additional questions regarding how the identified controls on the response to meltwater may be extended to the coupled climate system and consequently affect other components.
\\

Our investigation of different controls on the ocean response to meltwater is a non-exhaustive investigation of possible sensitivities of the response and aims to complement existing community efforts such as \cite{Swart2023TheDesign}. We presented results in a large perturbed parameter ensemble, which is only computationally feasible with simplifying assumptions, without aiming to investigate all questions surrounding meltwater injection (including magnitude, location, and best methods of prescription) nor all possible resolutions and parameters. The simplifications made should be explored more thoroughly in future work to understand the relevance of the results to similar experiments in full complexity coupled climate models. 
\\

One notable class of sensitivities that we do not fully examine is how meltwater is injected, as discussed in community efforts such as \cite{Swart2023TheDesign}. One uncertainty, recommended for investigation in Tier 3 of \cite{Swart2023TheDesign}, is whether to extract the latent heat from the ocean that was required to melt the ice, which has been suggested to substantially affect the heat budget \citep{Moorman2026TheMelt}. The extraction of latent heat was the recommended procedure in the \cite{Mathiot2017Explicit3.6} parameterization. Here, we simply injected the meltwater at 0 degrees, but the extraction of latent heat may change the dynamic responses found here, including the vertical plume that occurs in the deep meltwater perturbation case. We also neglected possible sensitivities to the zonal location of meltwater injection, instead opting to inject freshwater uniformly along the southern boundary due to the simplified domain. However, \cite{Moon2025AntarcticCirculation} showed that the zonal location of meltwater injection may be important, especially as the impact of meltwater on the abyssal cell is sensitive to the proximity of the meltwater injection to deep convection regions. 
\\

While our simple single-basin geometry allows for a large number of experiments, it ultimately results in omitting processes that may be important. Our simplified continental shelf does not have an Antarctic Slope Current, which has been demonstrated to trap meltwater on the continental shelf more effectively in models with a stronger current, whereas a weak current allows meltwater to enter the open ocean \citep{Beadling2022ImportanceChange}. The Antarctic slope current may depend on resolution \citep[e.g.,][]{Storkey2025ResolutionModels}, and thus be important for the sensitivity of the ocean response in a perturbed parameter ensemble which resolves it. Additionally, the importance of multiple basins, accurate basin widths, and coastline geometry for the large-scale ocean circulation has been demonstrated in previous studies \citep{Ragen2022TheCirculation,Talley2013ClosureOceans,Thompson2016ACirculation,Jones2016InterbasinCirculation,Newsom2018ReassessingCirculation,Sun2020TransientBasins}, and an implication of this is that the overturning cells in the present study may be altered when other basins are included.
\\

Finally, we highlight sensitivities associated with our perturbed parameter ensemble. The highest resolution utilized here was $\frac{1}{4}^{\circ}$, which is typically viewed as a ``grey zone" resolution such that the need for and correct way to implement mesoscale eddy parameterizations is unclear \citep{Hewitt2020ResolvingModels}. Examining higher resolution set-ups to more thoroughly test resolution dependence will strengthen our understanding of the impact of mesoscale eddies on the adjustment to Antarctic meltwater. This is particularly important as resolution can impact Southern Ocean stratification \citep[e.g.,][]{Marques2022NeverWorld2:Resolutions}, which we have demonstrated as strongly correlated to the sea level response and abyssal cell change. Additionally, the way we examined the impact of parametric uncertainty utilizing a one-at-a-time perturbed parameter ensemble, excludes non-linear relationships between different parameters which could amplify the spread of responses \citep[e.g.,][]{Eidhammer2024An6}. 
%We additionally stress that because ice sheets are not typically coupled to atmosphere-ocean models currently, there remain a range of open questions about the magnitude and location of realistic input, as well as uncertainty in the best methods of prescription, which need to be considered comprehensively to evaluate overall uncertainty in the system. Here, we sampled a limited subset of that uncertainty by utilizing multiple injection depths, but this should not be interpreted as attempting to capture the full range of questions surrounding meltwater injection. 
\\

Here, we have demonstrated that ocean model parameter values and resolution, as well as the depth of meltwater perturbations, influence sea level adjustment and the complex spatial and temporal structure of changes in circulation. We demonstrated these results in a simplified framework, and future work should further examine this in comprehensive coupled climate models, because the present results suggest that there are physical controls on the range of ocean responses to meltwater injection in climate model projections. These physical controls may be used to constrain and understand the ocean and climate response to anomalous Antarctic melt during the coming century.

\acknowledgments
This work was funded by NSF OCE grants 2048576 and 2048590 and supported by the New York University IT High Performance Computing resources, services, and staff expertise. This project is supported in part by the Eric and Wendy Schmidt AI in Science Postdoctoral Fellowship, a program of Schmidt Sciences. Without implying their endorsement, we thank Andy Thompson, Emma Beer, Pavel Perezhogin, and Sam Schulz for helpful discussions on this work.
%  Keep acknowledgments (note correct spelling: no ``e'' between the ``g'' and
% ``m'') as brief as possible. In general, acknowledge only direct help in
%  writing or research. Financial support (e.g., grant numbers) for the work done, 
%  for an author, or for the laboratory where the work was performed must be 
%  acknowledged here rather than as footnotes to the title or to an author's name.
%  Contribution numbers (if the work has been published by the author's institution 
%  or organization) should be placed in the acknowledgments rather than as 
%  footnotes to the title or to an author's name.

%%%%%%%%%%%%%%%%%%%%%%%%%%%%%%%%%%%%%%%%%%%%%%%%%%%%%%%%%%%%%%%%%%%%%
% DATA AVAILABILITY STATEMENT
%%%%%%%%%%%%%%%%%%%%%%%%%%%%%%%%%%%%%%%%%%%%%%%%%%%%%%%%%%%%%%%%%%%%%
% 
%
\datastatement 
By the time of publication, the simulation data and code to reproduce our analysis will be made available on Zenodo. 
%  The data availability statement is where authors should describe how the data underlying 
%  the findings within the article can be accessed and reused. Authors should attempt to 
%  provide unrestricted access to all data and materials underlying reported findings. 
%  If data access is restricted, authors must mention this in the statement. See
%  {http://www.ametsoc.org/PubsDataPolicy} for more info.

%%%%%%%%%%%%%%%%%%%%%%%%%%%%%%%%%%%%%%%%%%%%%%%%%%%%%%%%%%%%%%%%%%%%%
% APPENDIXES
%%%%%%%%%%%%%%%%%%%%%%%%%%%%%%%%%%%%%%%%%%%%%%%%%%%%%%%%%%%%%%%%%%%%%
%
%% If only one appendix, use

\appendix

%% If more than one appendix, use \appendix[<letter>], e.g.,
\appendix[A]
\appendixtitle{Dynamic sea level}
In the main text, we show the Northern Hemisphere dynamic sea level change at year 200 (see Figure \ref{fig:histogram}). As described in Section \ref{sec:methods}\ref{subsec:metrics}, the value of the dynamic sea level at year 200 with uncertainty is determined by a linear fit of the timeseries from year 100 to 200 where the adjustment is approximately linear. Here, we show the timeseries of the dynamic sea level for each experiment. 

\begin{figure*}[t]
\noindent\includegraphics[width=\textwidth,angle=0]{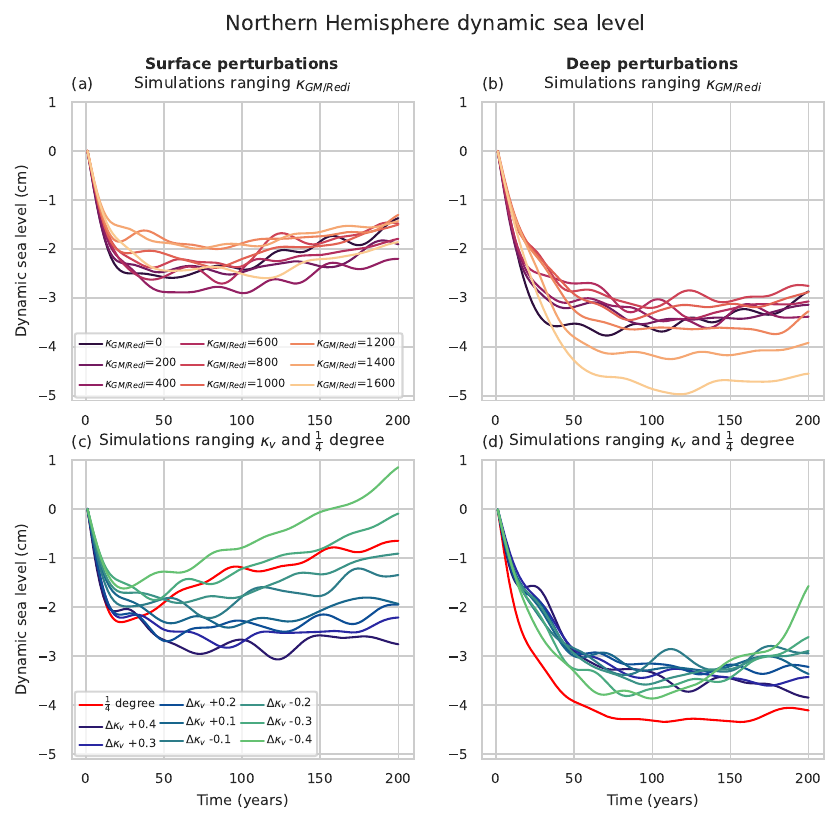}\\
\caption{Timeseries of Northern Hemisphere dynamic sea level in response to meltwater perturbations relative to control. The legend for panels (a) and (b) is shown on panel (a), while the legend for panels (c) and (d) is shown on panel (c). Panels (a) and (c) at the surface; (b) and (d) at depth. Curves are smoothed with a 6th order low-pass Butterworth filter with a cut-off period of 30 years. The top row (a and b) are for different $\kappa_{GM/Redi}$ parameters and the bottom row (c and d) are for varying $\kappa_v$. The $\frac{1}{4}^{\circ}$ results are shown in (c) and (d). An equivalent plot for the Southern Hemisphere dynamic sea level would be the exact negative of these results by definition.}  \label{fig:timeseries_dynamicsealevel}
%made in sealevel_plots

%  \caption{Timeseries of Northern Hemisphere dynamic sea level in each experiment relative to control for: (a) and (b)  surface meltwater input; (c) and (d) deep meltwater input. The top row (a and b) are for varying GM and the bottom row (c and d) are for different $\kappa_v$. The $\frac{1}{4}^{\circ}$ results are shown on all plots. The color for each experiment is shown in Figure \ref{fig:control_moc_setup}a.} \label{fig:timeseries_dynamicsealevel}
\end{figure*}
\begin{figure*}[t]
\noindent\includegraphics[width=\textwidth,angle=0]{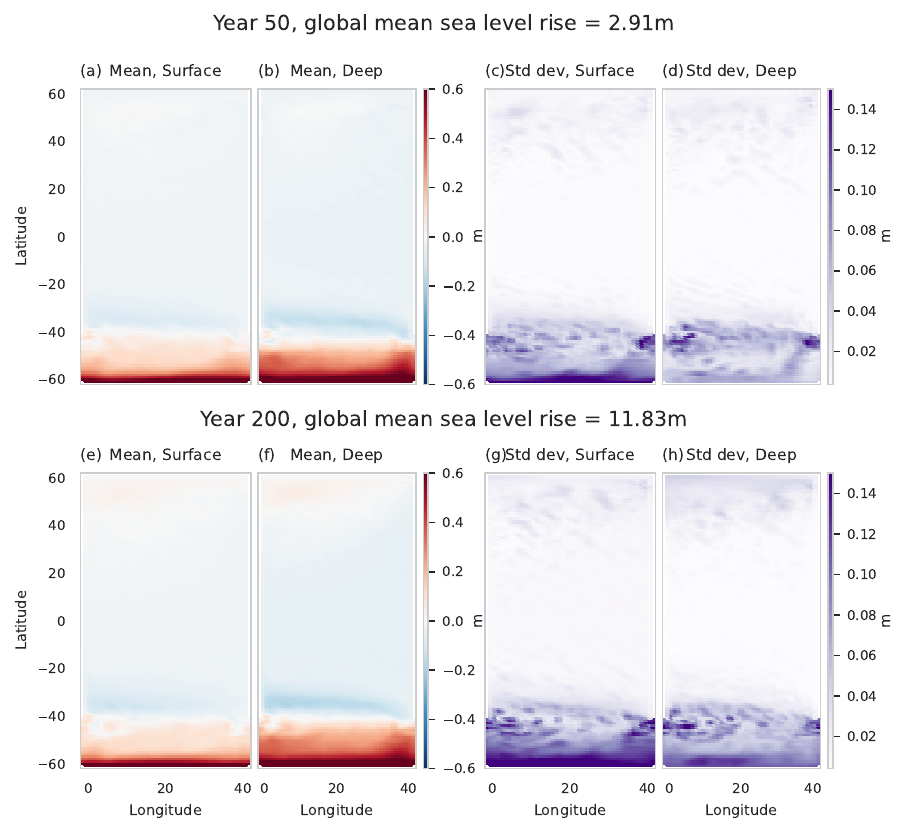}\\
\caption{The mean dynamic sea level pattern across model set-ups for each depth of perturbation, along with the standard deviation across model set-ups. Panels (a)-(d) after 50 years and panels (e)-(h) after 200 years.}  \label{fig:dynamicsealevelpattern}
\end{figure*}

\FloatBarrier
%\textcolor{gray}{Note for me, made in: dynamic\_sealevel\_timeseries}
%revisions made in sealevel_plots.ipynb in the analysis_2026_revisions folder
We also show the dynamic sea level pattern at years 50 and 200, along with the standard deviation across model set-ups in Figure \ref{fig:dynamicsealevelpattern}. 

\FloatBarrier
%made in more_sealevel

\appendix[B] 
\appendixtitle{Stratification in the control simulation}

\begin{figure*}[t]
\noindent\includegraphics[width=\textwidth,angle=0]{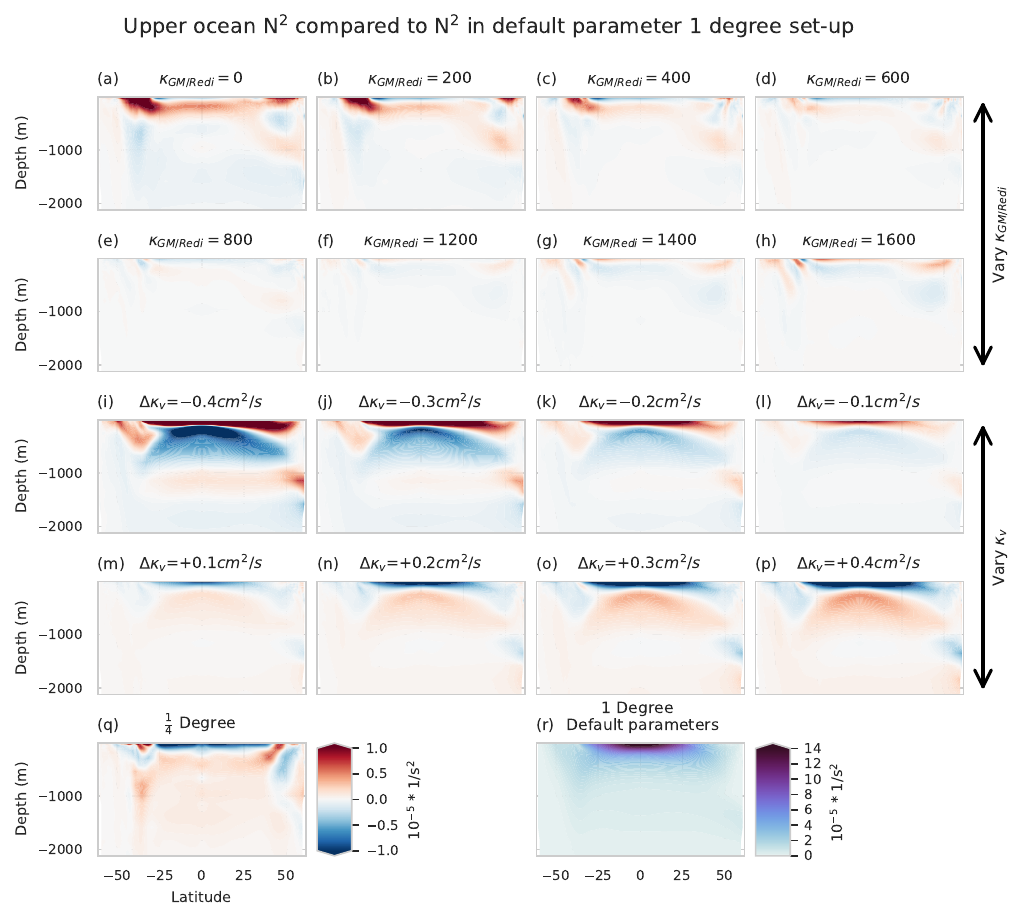}\\
  \caption{Control upper ocean $N^2$ in each case relative to the standard $1^{\circ}$ set-up: (a)-(h) with varying GM/Redi parameter ($\kappa_{\textrm{GM}}, \kappa_{\textrm{Redi}}$) as indicated in Figure \ref{fig:model_setup}b, with the parameter value used indicated in each subplot title; and (i)-(p) changing the vertical diffusivity as indicated in Figure \ref{fig:model_setup}d, with the value in the subfigure title indicating the shift from the default profile (uniformly over the whole depth); (q): the $\frac{1}{4}^{\circ}$ run; and (r): the 1-degree run with default parameter values (see Figure 1), i.e. the profile that is subtracted from all other set-ups.} \label{fig:control_n2}
\end{figure*}
\FloatBarrier
%made in streamfunctions_and_n2. Note didn't have to update for fixed runs cause from control but did anyway for a clean code cleanup

%\textcolor{gray}{Note for me, made in: appendix_figs_no18002000_reorder_may2025.ipynb} Update in appendix_figs_no18002000_reorder_aug2025.ipynb which is now in analysis_cleaned} 
%Bolus streamfunction (computed in depth coordinates) in the control runs. (a)-(g): the results from changing the GM/Redi parameter as indicated in Figure \ref{fig:model_setup}b. (h)-(o) the results from changing the vertical diffusivity as indicated in Figure \ref{fig:model_setup}c, the value in the subfigure title is the shift from the default profile (uniformly over the whole depth). (p): the $1^{\circ}$ run with default parameters. Note that the subplot labels are different than Figure \ref{fig:controlstreamfunction} because the $\frac{1}{4}^{\circ}$ run and the case of the GM parameter set to $0$ are not plotted due to not having bolus streamfunctions.

\appendix[C] 
\appendixtitle{Density change after perturbation}

\begin{figure*}[t]
\noindent\includegraphics[width=\textwidth,angle=0]{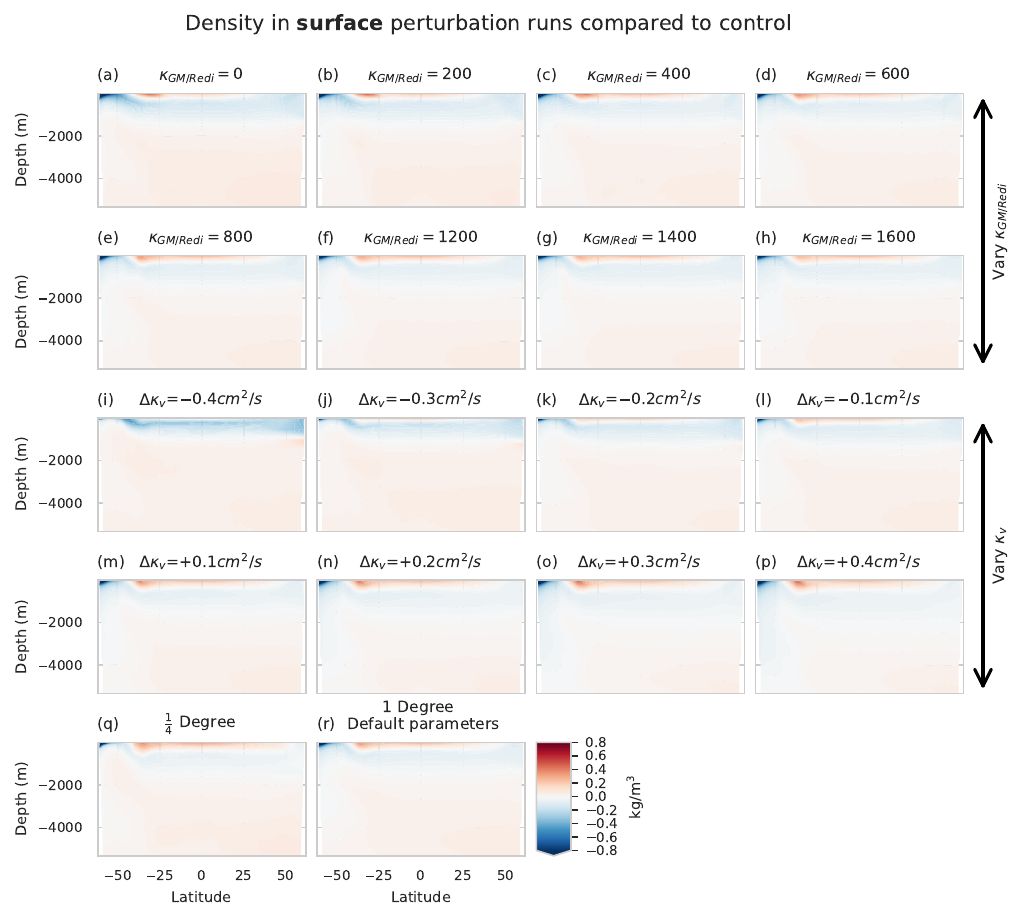}\\
\caption{Density change in a surface perturbation experiment relative to control: with varying GM/Redi parameter ($\kappa_{\textrm{GM}}, \kappa_{\textrm{Redi}}$) as indicated in Figure \ref{fig:model_setup}b, with the parameter value used indicated in each subplot title; and (i)-(p) changing the vertical diffusivity as indicated in Figure \ref{fig:model_setup}d, with the value in the subfigure title indicating the shift from the default profile (uniformly over the whole depth); (q): the $\frac{1}{4}^{\circ}$ run; and (r): the 1-degree run with default parameter values (see Figure \ref{fig:model_setup}). Note that for all plots, we subtract off the global mean change in density -- thus, this is the anomaly (compared to the global mean) of the density change.} \label{fig:density_surface}
\end{figure*}
\FloatBarrier
%made in plot_steric

\begin{figure*}[t]
\noindent\includegraphics[width=\textwidth,angle=0]{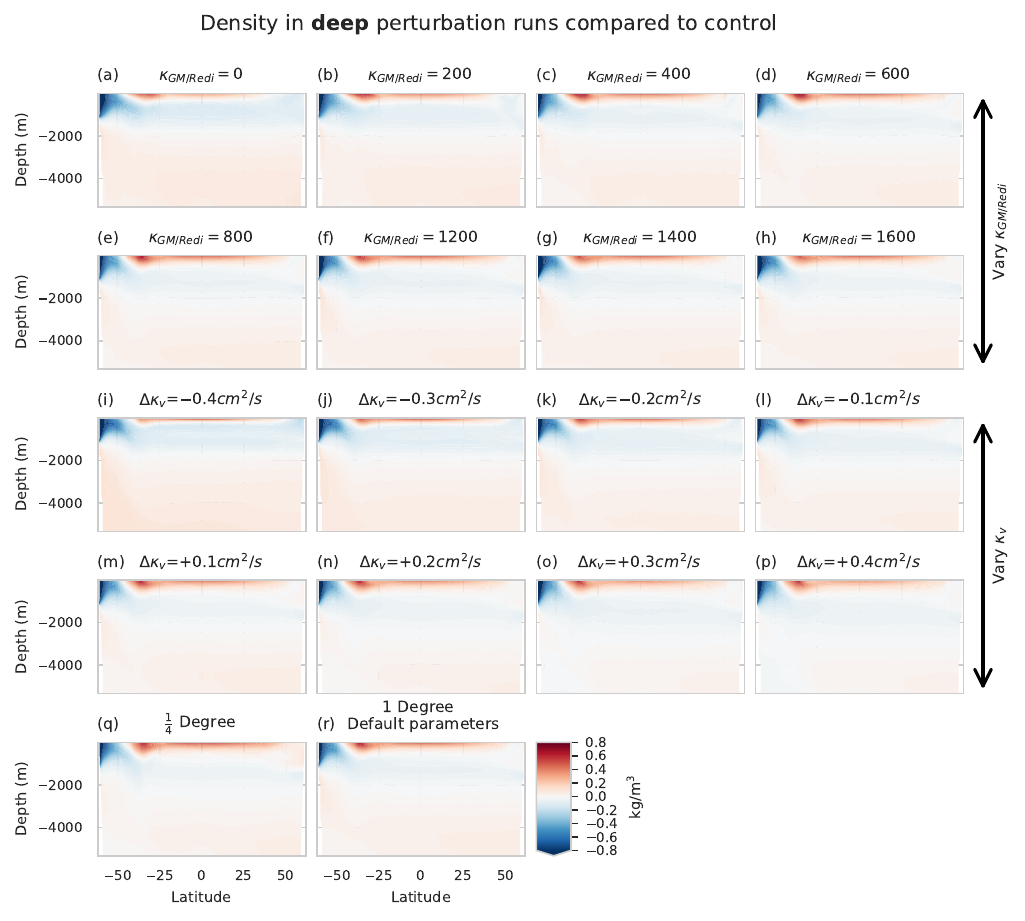} \caption{Density change in a deep perturbation experiment relative to control: with varying GM/Redi parameter ($\kappa_{\textrm{GM}}, \kappa_{\textrm{Redi}}$) as indicated in Figure \ref{fig:model_setup}b, with the parameter value used indicated in each subplot title; and (i)-(p) changing the vertical diffusivity as indicated in Figure \ref{fig:model_setup}d, with the value in the subfigure title indicating the shift from the default profile (uniformly over the whole depth); (q): the $\frac{1}{4}^{\circ}$ run; and (r): the 1-degree run with default parameter values (see Figure \ref{fig:model_setup}). Note that for all plots, we subtract off the global mean change in density -- thus, this is the anomaly (compared to the global mean) of the density change.} \label{fig:density_deep}
\end{figure*}
\FloatBarrier
\bibliographystyle{ametsocV6}
\bibliography{references_update}

@article{Mignot2006ADiffusivity,
    title = {{A Decomposition of the Atlantic Meridional Overturning Circulation into Physical Components Using Its Sensitivity to Vertical Diffusivity}},
    year = {2006},
    journal = {Journal of Physical Oceanography},
    author = {Mignot, Juliette and Levermann, Anders and Griesel, Alexa},
    number = {4},
    month = {4},
    pages = {636--650},
    volume = {36},
    publisher = {American Meteorological Society},
    url = {https://journals.ametsoc.org/view/journals/phoc/36/4/jpo2891.1.xml},
    doi = {10.1175/JPO2891.1},
    issn = {0022-3670}
}

@article{Marshall1997AComputers,
    title = {{A finite-volume, incompressible Navier Stokes model for studies of the ocean on parallel computers}},
    year = {1997},
    journal = {Journal of Geophysical Research: Oceans},
    author = {Marshall, John and Adcroft, Alistair and Hill, Chris and Perelman, Lev and Heisey, Curt},
    number = {C3},
    month = {3},
    pages = {5753--5766},
    volume = {102},
    publisher = {John Wiley {\&} Sons, Ltd},
    url = {https://onlinelibrary.wiley.com/doi/full/10.1029/96JC02775 https://onlinelibrary.wiley.com/doi/abs/10.1029/96JC02775 https://agupubs.onlinelibrary.wiley.com/doi/10.1029/96JC02775},
    doi = {10.1029/96JC02775},
    issn = {2156-2202}
}

@article{Gregory2005AConcentration,
    title = {{A model intercomparison of changes in the Atlantic thermohaline circulation in response to increasing atmospheric CO2 concentration}},
    year = {2005},
    journal = {Geophysical Research Letters},
    author = {Gregory, J. M. and Dixon, K. W. and Stouffer, R. J. and Weaver, A. J. and Driesschaert, E. and Eby, M. and Fichefet, T. and Hasumi, H. and Hu, A. and Jungclaus, J. H. and Kamenkovich, I. V. and Levermann, A. and Montoya, M. and Murakami, S. and Nawrath, S. and Oka, A. and Sokolov, A. P. and Thorpe, R. B.},
    number = {12},
    month = {6},
    pages = {},
    volume = {32},
    publisher = {John Wiley {\&} Sons, Ltd},
    url = {https://onlinelibrary.wiley.com/doi/full/10.1029/2005GL023209 https://onlinelibrary.wiley.com/doi/abs/10.1029/2005GL023209 https://agupubs.onlinelibrary.wiley.com/doi/10.1029/2005GL023209},
    isbn = {10.1029/2005},
    doi = {10.1029/2005GL023209},
    issn = {1944-8007}
}

@article{Thompson2016ACirculation,
    title = {{A Multibasin Residual-Mean Model for the Global Overturning Circulation}},
    year = {2016},
    journal = {Journal of Physical Oceanography},
    author = {Thompson, Andrew F. and Stewart, Andrew L. and Bischoff, Tobias},
    number = {9},
    month = {9},
    pages = {2583--2604},
    volume = {46},
    publisher = {American Meteorological Society},
    url = {https://journals.ametsoc.org/view/journals/phoc/46/9/jpo-d-15-0204.1.xml},
    doi = {10.1175/JPO-D-15-0204.1},
    issn = {0022-3670}
}

@article{Basinski-Ferris2025AEvolution,
    title = {{A Theory for How the Depth of Meltwater Injection Impacts Regional Sea Level Evolution}},
    year = {2025},
    journal = {Journal of Physical Oceanography},
    author = {Basinski-Ferris, Aurora and Zanna, Laure and Eisenman, Ian},
    number = {8},
    month = {8},
    pages = {1139--1154},
    volume = {55},
    publisher = {American Meteorological Society},
    url = {https://journals.ametsoc.org/view/journals/phoc/55/8/JPO-D-24-0153.1.xml},
    doi = {10.1175/JPO-D-24-0153.1},
    issn = {0022-3670}
}

@article{Nikurashin2011AOcean,
    title = {{A Theory of Deep Stratification and Overturning Circulation in the Ocean}},
    year = {2011},
    journal = {Journal of Physical Oceanography},
    author = {Nikurashin, Maxim and Vallis, Geoffrey},
    number = {3},
    month = {3},
    pages = {485--502},
    volume = {41},
    publisher = {American Meteorological Society},
    url = {https://journals.ametsoc.org/view/journals/phoc/41/3/2010jpo4529.1.xml},
    doi = {10.1175/2010JPO4529.1},
    issn = {0022-3670}
}

@article{Bryan1979AOcean,
    title = {{A water mass model of the World Ocean}},
    year = {1979},
    journal = {Journal of Geophysical Research: Oceans},
    author = {Bryan, K. and Lewis, L. J.},
    number = {C5},
    month = {5},
    pages = {2503--2517},
    volume = {84},
    publisher = {John Wiley {\&} Sons, Ltd},
    url = {https://onlinelibrary.wiley.com/doi/full/10.1029/JC084iC05p02503 https://onlinelibrary.wiley.com/doi/abs/10.1029/JC084iC05p02503 https://agupubs.onlinelibrary.wiley.com/doi/10.1029/JC084iC05p02503},
    doi = {10.1029/JC084IC05P02503},
    issn = {2156-2202}
}

@article{Li2023AbyssalMeltwater,
    title = {{Abyssal ocean overturning slowdown and warming driven by Antarctic meltwater}},
    year = {2023},
    journal = {Nature},
    author = {Li, Qian and England, Matthew H. and Hogg, Andrew Mc C. and Rintoul, Stephen R. and Morrison, Adele K.},
    number = {},
    month = {3},
    pages = {841--847},
    volume = {615},
    publisher = {Nature Publishing Group},
    url = {https://www.nature.com/articles/s41586-023-05762-w},
    doi = {10.1038/s41586-023-05762-w},
    issn = {1476-4687},
    pmid = {36991191}
}

@article{Eidhammer2024An6,
    title = {{An extensible perturbed parameter ensemble for the Community Atmosphere Model version 6}},
    year = {2024},
    journal = {Geoscientific Model Development},
    author = {Eidhammer, Trude and Gettelman, Andrew and Thayer-Calder, Katherine and Watson-Parris, Duncan and Elsaesser, Gregory and Morrison, Hugh and van Lier-Walqui, Marcus and Song, Ci and McCoy, Daniel},
    number = {21},
    month = {11},
    pages = {7835--7853},
    volume = {17},
    url = {https://gmd.copernicus.org/articles/17/7835/2024/},
    doi = {10.5194/gmd-17-7835-2024},
    issn = {1991-9603}
}

@article{Taylor2012AnDesign,
    title = {{An Overview of CMIP5 and the Experiment Design}},
    year = {2012},
    journal = {Bulletin of the American Meteorological Society},
    author = {Taylor, Karl E. and Stouffer, Ronald J. and Meehl, Gerald A.},
    number = {4},
    month = {4},
    pages = {485--498},
    volume = {93},
    publisher = {American Meteorological Society},
    url = {https://journals.ametsoc.org/view/journals/bams/93/4/bams-d-11-00094.1.xml},
    doi = {10.1175/BAMS-D-11-00094.1},
    issn = {00030007}
}

@article{Schmidt2023AnomalousForcing,
    title = {{Anomalous Meltwater From Ice Sheets and Ice Shelves Is a Historical Forcing}},
    year = {2023},
    journal = {Geophysical Research Letters},
    author = {Schmidt, Gavin A. and Romanou, Anastasia and Roach, Lettie A. and Mankoff, Kenneth D. and Li, Qian and Rye, Craig D. and Kelley, Maxwell and Marshall, John C. and Busecke, Julius J.M.},
    number = {24},
    month = {12},
    pages = {e2023GL106530},
    volume = {50},
    publisher = {John Wiley {\&} Sons, Ltd},
    url = {https://onlinelibrary.wiley.com/doi/full/10.1029/2023GL106530 https://onlinelibrary.wiley.com/doi/abs/10.1029/2023GL106530 https://agupubs.onlinelibrary.wiley.com/doi/10.1029/2023GL106530},
    doi = {10.1029/2023GL106530},
    issn = {1944-8007}
}

@article{Pritchard2012AntarcticShelves,
    title = {{Antarctic ice-sheet loss driven by basal melting of ice shelves}},
    year = {2012},
    journal = {Nature},
    author = {Pritchard, H. D. and Ligtenberg, S. R. M. and Fricker, H. A. and Vaughan, D. G. and Van Den Broeke, M. R. and Padman, L.},
    number = {},
    month = {4},
    pages = {502--505},
    volume = {484},
    publisher = {Nature Publishing Group},
    url = {https://www.nature.com/articles/nature10968},
    doi = {10.1038/nature10968},
    issn = {1476-4687}
}

@article{Dong2022AntarcticEffect,
    title = {{Antarctic Ice-Sheet Meltwater Reduces Transient Warming and Climate Sensitivity Through the Sea-Surface Temperature Pattern Effect}},
    year = {2022},
    journal = {Geophysical Research Letters},
    author = {Dong, Yue and Pauling, Andrew G. and Sadai, Shaina and Armour, Kyle C.},
    number = {24},
    month = {12},
    pages = {e2022GL101249},
    volume = {49},
    publisher = {John Wiley {\&} Sons, Ltd},
    url = {https://onlinelibrary.wiley.com/doi/full/10.1029/2022GL101249 https://onlinelibrary.wiley.com/doi/abs/10.1029/2022GL101249 https://agupubs.onlinelibrary.wiley.com/doi/10.1029/2022GL101249},
    doi = {10.1029/2022GL101249},
    issn = {1944-8007}
}

@article{An2024AntarcticTeleconnections,
    title = {{Antarctic meltwater reduces the Atlantic meridional overturning circulation through oceanic freshwater transport and atmospheric teleconnections}},
    year = {2024},
    journal = {Communications Earth {\&} Environment},
    author = {An, Soon-Il and Moon, Jun-Young and Dijkstra, Henk A. and Yang, Young-Min and Song, Hajoon},
    number = {},
    month = {9},
    pages = {},
    volume = {5},
    publisher = {Nature Publishing Group},
    url = {https://www.nature.com/articles/s43247-024-01670-7},
    doi = {10.1038/s43247-024-01670-7},
    issn = {2662-4435}
}

@article{Moon2025AntarcticCirculation,
    title = {{Antarctic meltwater spread pattern and its duration modulate abyssal circulation}},
    year = {2025},
    journal = {Communications Earth {\&} Environment},
    author = {Moon, Jun-Young and An, Soon-Il and Mandal, Gagan and Song, Hajoon and Yang, Young-Min and Park, So-Eun},
    number = {},
    month = {7},
    pages = {586},
    volume = {6},
    doi = {10.1038/s43247-025-02589-3},
    issn = {2662-4435}
}

@article{Newsom2023BackgroundEfficiency,
    title = {{Background Pycnocline Depth Constrains Future Ocean Heat Uptake Efficiency}},
    year = {2023},
    journal = {Geophysical Research Letters},
    author = {Newsom, Emily and Zanna, Laure and Gregory, Jonathan},
    number = {22},
    month = {11},
    pages = {e2023GL105673},
    volume = {50},
    publisher = {John Wiley {\&} Sons, Ltd},
    url = {https://onlinelibrary.wiley.com/doi/full/10.1029/2023GL105673 https://onlinelibrary.wiley.com/doi/abs/10.1029/2023GL105673 https://agupubs.onlinelibrary.wiley.com/doi/10.1029/2023GL105673},
    doi = {10.1029/2023GL105673},
    issn = {1944-8007},
    arxivId = {2307.11902}
}

@article{Depoorter2013CalvingShelves,
    title = {{Calving fluxes and basal melt rates of Antarctic ice shelves}},
    year = {2013},
    journal = {Nature},
    author = {Depoorter, M. A. and Bamber, J. L. and Griggs, J. A. and Lenaerts, J. T. M. and Ligtenberg, S. R. M. and Van Den Broeke, M. R. and Moholdt, G.},
    number = {},
    month = {9},
    pages = {89--92},
    volume = {502},
    publisher = {Nature Publishing Group},
    url = {https://www.nature.com/articles/nature12567},
    isbn = {2002201021},
    doi = {10.1038/nature12567},
    issn = {1476-4687},
    pmid = {24037377}
}

@article{Bronselaer2018ChangeMeltwater,
    title = {{Change in future climate due to Antarctic meltwater}},
    year = {2018},
    journal = {Nature},
    author = {Bronselaer, Ben and Winton, Michael and Griffies, Stephen M. and Hurlin, William J. and Rodgers, Keith B. and Sergienko, Olga V. and Stouffer, Ronald J. and Russell, Joellen L.},
    number = {},
    month = {11},
    pages = {53--58},
    volume = {564},
    publisher = {Nature Publishing Group},
    url = {https://www.nature.com/articles/s41586-018-0712-z},
    doi = {10.1038/s41586-018-0712-z},
    issn = {1476-4687},
    pmid = {30455421}
}

@article{Talley2013ClosureOceans,
    title = {{Closure of the global overturning circulation through the Indian, Pacific, and southern oceans}},
    year = {2013},
    journal = {Oceanography},
    author = {Talley, Lynne D.},
    number = {1},
    pages = {80--97},
    volume = {26},
    doi = {10.5670/OCEANOG.2013.07},
    issn = {10428275}
}

@article{Loose2023ComparingModel,
    title = {{Comparing Two Parameterizations for the Restratification Effect of Mesoscale Eddies in an Isopycnal Ocean Model}},
    year = {2023},
    journal = {Journal of Advances in Modeling Earth Systems},
    author = {Loose, Nora and Marques, Gustavo M. and Adcroft, Alistair and Bachman, Scott and Griffies, Stephen M. and Grooms, Ian and Hallberg, Robert W. and Jansen, Malte F.},
    number = {12},
    month = {12},
    pages = {e2022MS003518},
    volume = {15},
    publisher = {John Wiley {\&} Sons, Ltd},
    url = {https://onlinelibrary.wiley.com/doi/full/10.1029/2022MS003518 https://onlinelibrary.wiley.com/doi/abs/10.1029/2022MS003518 https://agupubs.onlinelibrary.wiley.com/doi/10.1029/2022MS003518},
    doi = {10.1029/2022MS003518},
    issn = {1942-2466}
}

@article{Nayak2024ControlsModels,
    title = {{Controls on the Strength and Structure of the Atlantic Meridional Overturning Circulation in Climate Models}},
    year = {2024},
    journal = {Geophysical Research Letters},
    author = {Nayak, Manali S. and Bonan, David B. and Newsom, Emily R. and Thompson, Andrew F.},
    number = {11},
    month = {6},
    pages = {e2024GL109055},
    volume = {51},
    publisher = {John Wiley {\&} Sons, Ltd},
    url = {https://onlinelibrary.wiley.com/doi/full/10.1029/2024GL109055 https://onlinelibrary.wiley.com/doi/abs/10.1029/2024GL109055 https://agupubs.onlinelibrary.wiley.com/doi/10.1029/2024GL109055},
    doi = {10.1029/2024GL109055},
    issn = {1944-8007}
}

@article{Oka2025DeepEstimates,
    title = {{Deep ocean mixing mismatch between model and observational estimates}},
    year = {2025},
    journal = {Communications Earth {\&} Environment},
    author = {Oka, Akira},
    number = {},
    month = {2},
    pages = {108},
    volume = {6},
    publisher = {Nature Publishing Group},
    url = {https://www.nature.com/articles/s43247-025-02027-4},
    doi = {10.1038/s43247-025-02027-4},
    issn = {2662-4435}
}

@article{Huber2017DriversUptake,
    title = {{Drivers of uncertainty in simulated ocean circulation and heat uptake}},
    year = {2017},
    journal = {Geophysical Research Letters},
    author = {Huber, Markus B. and Zanna, Laure},
    number = {3},
    month = {2},
    pages = {1402--1413},
    volume = {44},
    publisher = {Blackwell Publishing Ltd},
    doi = {10.1002/2016GL071587},
    issn = {19448007}
}

@article{Munday2013EddyCurrents,
    title = {{Eddy Saturation of Equilibrated Circumpolar Currents}},
    year = {2013},
    journal = {Journal of Physical Oceanography},
    author = {Munday, David R. and Johnson, Helen L. and Marshall, David P.},
    number = {3},
    month = {3},
    pages = {507--532},
    volume = {43},
    publisher = {American Meteorological Society},
    url = {https://journals.ametsoc.org/view/journals/phoc/43/3/jpo-d-12-095.1.xml},
    doi = {10.1175/JPO-D-12-095.1},
    issn = {0022-3670}
}

@article{Mathiot2017Explicit3.6,
    title = {{Explicit representation and parametrised impacts of under ice shelf seas in the z * coordinate ocean model NEMO 3.6}},
    year = {2017},
    journal = {Geosci. Model Dev},
    author = {Mathiot, Pierre and Jenkins, Adrian and Harris, Christopher and Madec, Gurvan},
    pages = {2849--2874},
    volume = {10},
    url = {https://doi.org/10.5194/gmd-10-2849-2017},
    doi = {10.5194/gmd-10-2849-2017}
}

@article{Sadai2020FutureWarming,
    title = {{Future climate response to Antarctic Ice Sheet melt caused by anthropogenic warming}},
    year = {2020},
    journal = {Science Advances},
    author = {Sadai, Shaina and Condron, Alan and DeConto, Robert and Pollard, David},
    number = {39},
    month = {9},
    volume = {6},
    publisher = {American Association for the Advancement of Science},
    url = {https://www.science.org/doi/10.1126/sciadv.aaz1169},
    doi = {10.1126/sciadv.aaz1169},
    issn = {23752548},
    pmid = {32967838}
}

@article{Park2023FutureModel,
    title = {{Future sea-level projections with a coupled atmosphere-ocean-ice-sheet model}},
    year = {2023},
    journal = {Nature Communications},
    author = {Park, Jun Young and Schloesser, Fabian and Timmermann, Axel and Choudhury, Dipayan and Lee, June Yi and Nellikkattil, Arjun Babu},
    number = {},
    month = {2},
    pages = {636},
    volume = {14},
    publisher = {Nature Publishing Group},
    url = {https://www.nature.com/articles/s41467-023-36051-9},
    doi = {10.1038/s41467-023-36051-9},
    issn = {2041-1723},
    pmid = {36788205}
}

@article{Li2023GlobalStudy,
    title = {{Global Climate Impacts of Greenland and Antarctic Meltwater: A Comparative Study}},
    year = {2023},
    journal = {Journal of Climate},
    author = {Li, Qian and Marshall, John and Rye, Craig D. and Romanou, Anastasia and Rind, David and Kelley, Maxwell},
    number = {11},
    month = {5},
    pages = {3571--3590},
    volume = {36},
    publisher = {American Meteorological Society},
    url = {https://journals.ametsoc.org/view/journals/clim/36/11/JCLI-D-22-0433.1.xml},
    doi = {10.1175/JCLI-D-22-0433.1},
    issn = {0894-8755}
}

@article{Golledge2019GlobalMelt,
    title = {{Global environmental consequences of twenty-first-century ice-sheet melt}},
    year = {2019},
    journal = {Nature},
    author = {Golledge, Nicholas R. and Keller, Elizabeth D. and Gomez, Natalya and Naughten, Kaitlin A. and Bernales, Jorge and Trusel, Luke D. and Edwards, Tamsin L.},
    number = {},
    month = {2},
    pages = {65--72},
    volume = {566},
    publisher = {Nature Publishing Group},
    url = {https://www.nature.com/articles/s41586-019-0889-9},
    doi = {10.1038/s41586-019-0889-9},
    issn = {1476-4687},
    pmid = {30728520}
}

@article{Ma2011GlobalOcean,
    title = {{Global Teleconnections in Response to Freshening over the Antarctic Ocean}},
    year = {2011},
    journal = {Journal of Climate},
    author = {Ma, Hao and Wu, Lixin},
    number = {4},
    month = {2},
    pages = {1071--1088},
    volume = {24},
    publisher = {American Meteorological Society},
    url = {https://journals.ametsoc.org/view/journals/clim/24/4/2010jcli3634.1.xml},
    doi = {10.1175/2010JCLI3634.1},
    issn = {0894-8755}
}

@article{Winton2014HasSensitivity,
    title = {{Has coarse ocean resolution biased simulations of transient climate sensitivity?}},
    year = {2014},
    journal = {Geophysical Research Letters},
    author = {Winton, Michael and Anderson, Whit G. and Delworth, Thomas L. and Griffies, Stephen M. and Hurlin, William J. and Rosati, Anthony},
    number = {23},
    month = {12},
    pages = {8522--8529},
    volume = {41},
    publisher = {Blackwell Publishing Ltd},
    doi = {10.1002/2014GL061523},
    issn = {19448007}
}

@article{Rignot2013Ice-shelfAntarctica,
    title = {{Ice-shelf melting around Antarctica}},
    year = {2013},
    journal = {Science},
    author = {Rignot, E. and Jacobs, S. and Mouginot, J. and Scheuchl, B.},
    number = {6143},
    month = {7},
    pages = {266--270},
    volume = {341},
    publisher = {American Association for the Advancement of Science},
    url = {https://www.science.org/doi/10.1126/science.1235798},
    doi = {10.1126/science.1235798},
    issn = {10959203}
}

@article{Saenko2018ImpactModel,
    title = {{Impact of Mesoscale Eddy Transfer on Heat Uptake in an Eddy-Parameterizing Ocean Model}},
    year = {2018},
    journal = {Journal of Climate},
    author = {Saenko, Oleg A. and Yang, Duo and Gregory, Jonathan M.},
    number = {20},
    month = {10},
    pages = {8589--8606},
    volume = {31},
    publisher = {American Meteorological Society},
    url = {https://journals.ametsoc.org/view/journals/clim/31/20/jcli-d-18-0186.1.xml},
    doi = {10.1175/JCLI-D-18-0186.1},
    issn = {0894-8755}
}

@article{Beadling2022ImportanceChange,
    title = {{Importance of the Antarctic Slope Current in the Southern Ocean Response to Ice Sheet Melt and Wind Stress Change}},
    year = {2022},
    journal = {Journal of Geophysical Research: Oceans},
    author = {Beadling, R. L. and Krasting, J. P. and Griffies, S. M. and Hurlin, W. J. and Bronselaer, B. and Russell, J. L. and MacGilchrist, G. A. and Tesdal, J. E. and Winton, M.},
    number = {5},
    month = {5},
    pages = {e2021JC017608},
    volume = {127},
    publisher = {John Wiley {\&} Sons, Ltd},
    url = {https://onlinelibrary.wiley.com/doi/full/10.1029/2021JC017608 https://onlinelibrary.wiley.com/doi/abs/10.1029/2021JC017608 https://agupubs.onlinelibrary.wiley.com/doi/10.1029/2021JC017608},
    isbn = {10.1029/2021},
    doi = {10.1029/2021JC017608},
    issn = {2169-9291}
}

@article{Bronselaer2020ImportanceOcean,
    title = {{Importance of wind and meltwater for observed chemical and physical changes in the Southern Ocean}},
    year = {2020},
    journal = {Nature Geoscience},
    author = {Bronselaer, Ben and Russell, Joellen L. and Winton, Michael and Williams, Nancy L. and Key, Robert M. and Dunne, John P. and Feely, Richard A. and Johnson, Kenneth S. and Sarmiento, Jorge L.},
    number = {},
    month = {1},
    pages = {35--42},
    volume = {13},
    publisher = {Nature Publishing Group},
    url = {https://www.nature.com/articles/s41561-019-0502-8},
    doi = {10.1038/s41561-019-0502-8},
    issn = {1752-0908}
}

@article{Zika2018ImprovedWarming,
    title = {{Improved estimates of water cycle change from ocean salinity: the key role of ocean warming}},
    year = {2018},
    journal = {Environmental Research Letters},
    author = {Zika, Jan D and Skliris, Nikolaos and Blaker, Adam T and Marsh, Robert and Nurser, A J George and Joser, Simon A},
    number = {7},
    month = {7},
    pages = {074036},
    volume = {13},
    publisher = {IOP Publishing},
    url = {http://iopscience.iop.org/article/10.1088/1748-9326/10/9/094021/meta},
    doi = {10.1088/1748-9326/aace42}
}

@article{Jones2016InterbasinCirculation,
    title = {{Interbasin Transport of the Meridional Overturning Circulation}},
    year = {2016},
    journal = {Journal of Physical Oceanography},
    author = {Jones, C. S. and Cessi, Paola},
    number = {4},
    month = {4},
    pages = {1157--1169},
    volume = {46},
    publisher = {American Meteorological Society},
    url = {https://journals.ametsoc.org/view/journals/phoc/46/4/jpo-d-15-0197.1.xml},
    doi = {10.1175/JPO-D-15-0197.1},
    issn = {0022-3670}
}

@article{Gent1990IsopycnalModels,
    title = {{Isopycnal Mixing in Ocean Circulation Models}},
    year = {1990},
    journal = {Journal of Physical Oceanography},
    author = {Gent, Peter R. and McWilliams, James C.},
    number = {1},
    month = {1},
    pages = {150--155},
    volume = {20},
    publisher = {American Meteorological Society},
    url = {https://journals.ametsoc.org/view/journals/phoc/20/1/1520-0485_1990_020_0150_imiocm_2_0_co_2.xml},
    doi = {10.1175/1520-0485(1990)020<0150:IMIOCM>2.0.CO;2},
    issn = {0022-3670}
}

@article{Otosaka2023Mass2020,
    title = {{Mass balance of the Greenland and Antarctic ice sheets from 1992 to 2020}},
    year = {2023},
    journal = {Earth System Science Data},
    author = {Otosaka, Inès N. and Shepherd, Andrew and Ivins, Erik R. and Schlegel, Nicole Jeanne and Amory, Charles and Van Den Broeke, Michiel R. and Horwath, Martin and Joughin, Ian and King, Michalea D. and Krinner, Gerhard and Nowicki, Sophie and Payne, Anthony J. and Rignot, Eric and Scambos, Ted and Simon, Karen M. and Smith, Benjamin E. and S{\o}rensen, Louise S. and Velicogna, Isabella and Whitehouse, Pippa L. and Geruo, A. and Agosta, Cécile and Ahlstr{\o}m, Andreas P. and Blazquez, Alejandro and Colgan, William and Engdahl, Marcus E. and Fettweis, Xavier and Forsberg, Rene and Gall{\'{e}}e, Hubert and Gardner, Alex and Gilbert, Lin and Gourmelen, Noel and Groh, Andreas and Gunter, Brian C. and Harig, Christopher and Helm, Veit and Khan, Shfaqat Abbas and Kittel, Christoph and Konrad, Hannes and Langen, Peter L. and Lecavalier, Benoit S. and Liang, Chia Chun and Loomis, Bryant D. and McMillan, Malcolm and Melini, Daniele and Mernild, Sebastian H. and Mottram, Ruth and Mouginot, Jeremie and Nilsson, Johan and No{\"{e}}l, Brice and Pattle, Mark E. and Peltier, William R. and Pie, Nadege and Roca, Mònica and Sasgen, Ingo and Save, Himanshu V. and Seo, Ki Weon and Scheuchl, Bernd and Schrama, Ernst J.O. and Schr{\"{o}}der, Ludwig and Simonsen, Sebastian B. and Slater, Thomas and Spada, Giorgio and Sutterley, Tyler C. and Vishwakarma, Bramha Dutt and Van Wessem, Jan Melchior and Wiese, David and Van Der Wal, Wouter and Wouters, Bert},
    number = {4},
    month = {4},
    pages = {1597--1616},
    volume = {15},
    publisher = {Copernicus Publications},
    doi = {10.5194/ESSD-15-1597-2023},
    issn = {18663516}
}

@article{Jackett1995MinimalStability,
    title = {{Minimal Adjustment of Hydrographic Profiles to Achieve Static Stability}},
    year = {1995},
    journal = {Journal of Atmospheric and Oceanic Technology},
    author = {Jackett, David R. and Mcdougall, Trevor J.},
    number = {2},
    month = {4},
    pages = {381--389},
    volume = {12},
    publisher = {American Meteorological Society},
    url = {http://journals.ametsoc.org/doi/10.1175/1520-0426(1995)012<0381:MAOHPT>2.0.CO;2},
    doi = {10.1175/1520-0426(1995)012<0381:MAOHPT>2.0.CO;2},
    issn = {0739-0572}
}

@article{Marques2022NeverWorld2:Resolutions,
    title = {{NeverWorld2: An idealized model hierarchy to investigate ocean mesoscale eddies across resolutions}},
    year = {2022},
    journal = {Geoscientific Model Development},
    author = {Marques, Gustavo M. and Loose, Nora and Yankovsky, Elizabeth and Steinberg, Jacob M. and Chang, Chiung Yin and Bhamidipati, Neeraja and Adcroft, Alistair and Fox-Kemper, Baylor and Griffies, Stephen M. and Hallberg, Robert W. and Jansen, Malte F. and Khatri, Hemant and Zanna, Laure},
    number = {17},
    month = {9},
    pages = {6567--6579},
    volume = {15},
    publisher = {Copernicus Publications},
    doi = {10.5194/GMD-15-6567-2022},
    issn = {19919603}
}

@article{Bonan2025ObservationalWeakening,
    title = {{Observational constraints imply limited future Atlantic meridional overturning circulation weakening}},
    year = {2025},
    journal = {Nature Geoscience},
    author = {Bonan, David B. and Thompson, Andrew F. and Schneider, Tapio and Zanna, Laure and Armour, Kyle C. and Sun, Shantong},
    number = {},
    month = {6},
    pages = {479--487},
    volume = {18},
    url = {https://www.nature.com/articles/s41561-025-01709-0},
    doi = {10.1038/s41561-025-01709-0},
    issn = {1752-0894}
}

@article{Redi1982OceanicRotation,
    title = {{Oceanic Isopycnal Mixing by Coordinate Rotation}},
    year = {1982},
    journal = {Journal of Physical Oceanography},
    author = {Redi, Martha H.},
    number = {10},
    month = {10},
    pages = {1154--1158},
    volume = {12},
    publisher = {American Meteorological Society},
    url = {https://journals.ametsoc.org/view/journals/phoc/12/10/1520-0485_1982_012_1154_oimbcr_2_0_co_2.xml},
    doi = {10.1175/1520-0485(1982)012<1154:OIMBCR>2.0.CO;2},
    issn = {0022-3670}
}

@article{Todd2020OceanOnlyChange,
    title = {{Ocean‐Only FAFMIP: Understanding Regional Patterns of Ocean Heat Content and Dynamic Sea Level Change}},
    year = {2020},
    journal = {Journal of Advances in Modeling Earth Systems},
    author = {Todd, Alexander and Zanna, Laure and Couldrey, Matthew and Gregory, Jonathan and Wu, Quran and Church, John A. and Farneti, Riccardo and Navarro‐Labastida, René and Lyu, Kewei and Saenko, Oleg and Yang, Duo and Zhang, Xuebin},
    number = {8},
    month = {8},
    pages = {e2019MS002027},
    volume = {12},
    publisher = {Blackwell Publishing Ltd},
    url = {https://onlinelibrary.wiley.com/doi/10.1029/2019MS002027},
    doi = {10.1029/2019MS002027},
    issn = {1942-2466}
}

@article{Farrell1976OnLevel,
    title = {{On Postglacial Sea Level}},
    year = {1976},
    journal = {Geophysical Journal International},
    author = {Farrell, W. E. and Clark, J. A.},
    number = {3},
    month = {9},
    pages = {647--667},
    volume = {46},
    publisher = {Oxford Academic},
    url = {https://dx.doi.org/10.1111/j.1365-246X.1976.tb01252.x},
    doi = {10.1111/J.1365-246X.1976.TB01252.X},
    issn = {0956-540X}
}

@article{Stewart2014OnOcean,
    title = {{On the Importance of Surface Forcing in Conceptual Models of the Deep Ocean}},
    year = {2014},
    journal = {Journal of Physical Oceanography},
    author = {Stewart, Andrew L. and Ferrari, Raffaele and Thompson, Andrew F.},
    number = {3},
    month = {3},
    pages = {891--899},
    volume = {44},
    publisher = {American Meteorological Society},
    url = {https://journals.ametsoc.org/view/journals/phoc/44/3/jpo-d-13-0206.1.xml},
    doi = {10.1175/JPO-D-13-0206.1},
    issn = {0022-3670}
}

@article{Arnscheidt2021OnShelves,
    title = {{On the Settling Depth of Meltwater Escaping from beneath Antarctic Ice Shelves}},
    year = {2021},
    journal = {Journal of Physical Oceanography},
    author = {Arnscheidt, Constantin W. and Marshall, John and Dutrieux, Pierre and Rye, Craig D. and Ramadhan, Ali},
    number = {7},
    month = {7},
    pages = {2257--2270},
    volume = {51},
    publisher = {American Meteorological Society},
    url = {https://journals.ametsoc.org/view/journals/phoc/51/7/JPO-D-20-0286.1.xml},
    doi = {10.1175/JPO-D-20-0286.1},
    issn = {0022-3670}
}

@article{Eyring2016OverviewOrganization,
    title = {{Overview of the Coupled Model Intercomparison Project Phase 6 (CMIP6) experimental design and organization}},
    year = {2016},
    journal = {Geoscientific Model Development},
    author = {Eyring, Veronika and Bony, Sandrine and Meehl, Gerald A. and Senior, Catherine A. and Stevens, Bjorn and Stouffer, Ronald J. and Taylor, Karl E.},
    number = {5},
    month = {5},
    pages = {1937--1958},
    volume = {9},
    publisher = {Copernicus GmbH},
    doi = {10.5194/GMD-9-1937-2016},
    issn = {19919603}
}

@article{Lago2019ProjectedContributions,
    title = {{Projected Slowdown of Antarctic Bottom Water Formation in Response to Amplified Meltwater Contributions}},
    year = {2019},
    journal = {Journal of Climate},
    author = {Lago, Véronique and England, Matthew H.},
    number = {19},
    month = {10},
    pages = {6319--6335},
    volume = {32},
    publisher = {American Meteorological Society},
    url = {https://journals.ametsoc.org/view/journals/clim/32/19/jcli-d-18-0622.1.xml},
    doi = {10.1175/JCLI-D-18-0622.1},
    issn = {0894-8755}
}

@article{Mitrovica2018QuantifyingFlux,
    title = {{Quantifying the Sensitivity of Sea Level Change in Coastal Localities to the Geometry of Polar Ice Mass Flux}},
    year = {2018},
    journal = {Journal of Climate},
    author = {Mitrovica, Jerry X. and Hay, Carling C. and Kopp, Robert E. and Harig, Christopher and Latychev, Konstantin},
    number = {9},
    month = {5},
    pages = {3701--3709},
    volume = {31},
    publisher = {American Meteorological Society},
    url = {https://journals.ametsoc.org/view/journals/clim/31/9/jcli-d-17-0465.1.xml},
    doi = {10.1175/JCLI-D-17-0465.1},
    issn = {0894-8755}
}

@article{Lorbacher2012RapidMelting,
    title = {{Rapid barotropic sea level rise from ice sheet melting}},
    year = {2012},
    journal = {Journal of Geophysical Research: Oceans},
    author = {Lorbacher, K. and Marsland, S. J. and Church, J. A. and Griffies, S. M. and Stammer, D.},
    number = {C6},
    month = {6},
    pages = {},
    volume = {117},
    publisher = {John Wiley {\&} Sons, Ltd},
    url = {https://onlinelibrary.wiley.com/doi/full/10.1029/2011JC007733 https://onlinelibrary.wiley.com/doi/abs/10.1029/2011JC007733 https://agupubs.onlinelibrary.wiley.com/doi/10.1029/2011JC007733},
    doi = {10.1029/2011JC007733},
    issn = {2156-2202}
}

@article{Newsom2018ReassessingCirculation,
    title = {{Reassessing the Role of the Indo-Pacific in the Ocean's Global Overturning Circulation}},
    year = {2018},
    journal = {Geophysical Research Letters},
    author = {Newsom, Emily R. and Thompson, Andrew F.},
    number = {22},
    month = {11},
    pages = {12422--12431},
    volume = {45},
    publisher = {John Wiley {\&} Sons, Ltd},
    url = {https://onlinelibrary.wiley.com/doi/full/10.1029/2018GL080350 https://onlinelibrary.wiley.com/doi/abs/10.1029/2018GL080350 https://agupubs.onlinelibrary.wiley.com/doi/10.1029/2018GL080350},
    doi = {10.1029/2018GL080350},
    issn = {1944-8007}
}

@article{Storkey2025ResolutionModels,
    title = {{Resolution dependence of interlinked Southern Ocean biases in global coupled HadGEM3 models}},
    year = {2025},
    journal = {Geoscientific Model Development},
    author = {Storkey, David and Mathiot, Pierre and Bell, Michael J. and Copsey, Dan and Guiavarc'H, Catherine and Hewitt, Helene T. and Ridley, Jeff and Roberts, Malcolm J.},
    number = {9},
    month = {5},
    pages = {2725--2745},
    volume = {18},
    publisher = {Copernicus Publications},
    doi = {10.5194/GMD-18-2725-2025},
    issn = {19919603}
}

@article{Hewitt2020ResolvingModels,
    title = {{Resolving and Parameterising the Ocean Mesoscale in Earth System Models}},
    year = {2020},
    journal = {Current Climate Change Reports},
    author = {Hewitt, Helene T. and Roberts, Malcolm and Mathiot, Pierre and Biastoch, Arne and Blockley, Ed and Chassignet, Eric P. and Fox-Kemper, Baylor and Hyder, Pat and Marshall, David P. and Popova, Ekaterina and Treguier, Anne-Marie and Zanna, Laure and Yool, Andrew and Yu, Yongqiang and Beadling, Rebecca and Bell, Mike and Kuhlbrodt, Till and Arsouze, Thomas and Bellucci, Alessio and Castruccio, Fred and Gan, Bolan and Putrasahan, Dian and Roberts, Christopher D. and Van Roekel, Luke and Zhang, Qiuying},
    number = {},
    month = {12},
    pages = {137--152},
    volume = {6},
    publisher = {Springer},
    url = {https://link.springer.com/10.1007/s40641-020-00164-w},
    doi = {10.1007/s40641-020-00164-w},
    issn = {2198-6061}
}

@article{Stammer2008ResponseMelting,
    title = {{Response of the global ocean to Greenland and Antarctic ice melting}},
    year = {2008},
    journal = {Journal of Geophysical Research: Oceans},
    author = {Stammer, D.},
    number = {C6},
    month = {6},
    volume = {113},
    publisher = {John Wiley {\&} Sons, Ltd},
    url = {https://onlinelibrary.wiley.com/doi/full/10.1029/2006JC004079 https://onlinelibrary.wiley.com/doi/abs/10.1029/2006JC004079 https://agupubs.onlinelibrary.wiley.com/doi/10.1029/2006JC004079},
    doi = {10.1029/2006JC004079},
    issn = {2156-2202}
}

@article{Luongo2024RetainingSimulations,
    title = {{Retaining Short-Term Variability Reduces Mean State Biases in Wind Stress Overriding Simulations}},
    year = {2024},
    journal = {Journal of Advances in Modeling Earth Systems},
    author = {Luongo, Matthew T. and Brizuela, Noel G. and Eisenman, Ian and Xie, Shang Ping},
    number = {2},
    month = {2},
    pages = {e2023MS003665},
    volume = {16},
    publisher = {John Wiley {\&} Sons, Ltd},
    url = {https://onlinelibrary.wiley.com/doi/full/10.1029/2023MS003665 https://onlinelibrary.wiley.com/doi/abs/10.1029/2023MS003665 https://agupubs.onlinelibrary.wiley.com/doi/10.1029/2023MS003665},
    doi = {10.1029/2023MS003665},
    issn = {1942-2466}
}

@incollection{Oppenheimer2019SeaCommunities,
    title = {{Sea Level Rise and Implications for Low-Lying Islands, Coasts and Communities}},
    year = {2019},
    booktitle = {IPCC Special Report on the Ocean and Cryosphere in a Changing Climate},
    author = {Oppenheimer, M and Glavovic, B C and Hinkel, J and van de Wal, R and Magnan, A K and Abd-Elgawad, A and Cai, R and Cifuentes-Jara, M and DeConto, R M and Ghosh, T and Hay, J and Isla, F and Marzeion, B and Meyssignac, B and Sebesvari, Z},
    editor = {P{\"{o}}rtner, H.-O. and Roberts, D C and Masson-Delmotte, V and Zhai, P and Tignor, M and Poloczanska, E and Mintenbeck, K and Alegr{\'{i}}a, A and Nicolai, M and Okem, A and Petzold, J and Rama, B and Weyer, N M},
    pages = {321--445},
    publisher = {Cambridge University Press},
    doi = {10.1017/9781009157964.006}
}

@article{Armour2024Sea-surfaceSensitivity,
    title = {{Sea-surface temperature pattern effects have slowed global warming and biased warming-based constraints on climate sensitivity}},
    year = {2024},
    journal = {Proceedings of the National Academy of Sciences},
    author = {Armour, Kyle C. and Proistosescu, Cristian and Dong, Yue and Hahn, Lily C. and Blanchard-Wrigglesworth, Edward and Pauling, Andrew G. and Jnglin Wills, Robert C. and Andrews, Timothy and Stuecker, Malte F. and Po-Chedley, Stephen and Mitevski, Ivan and Forster, Piers M. and Gregory, Jonathan M.},
    number = {12},
    month = {3},
    pages = {e2312093121},
    volume = {121},
    publisher = {National Academy of Sciences},
    url = {https://pnas.org/doi/10.1073/pnas.2312093121},
    doi = {10.1073/pnas.2312093121},
    issn = {0027-8424}
}

@article{Wickramage2023SensitivityResolution,
    title = {{Sensitivity of MPI-ESM Sea Level Projections to Its Ocean Spatial Resolution}},
    year = {2023},
    journal = {Journal of Climate},
    author = {Wickramage, Chathurika and K{\"{o}}hl, Armin and Jungclaus, Johann and Stammer, Detlef},
    number = {6},
    month = {2},
    pages = {1957--1980},
    volume = {36},
    publisher = {American Meteorological Society},
    url = {https://journals.ametsoc.org/view/journals/clim/36/6/JCLI-D-22-0418.1.xml},
    doi = {10.1175/JCLI-D-22-0418.1},
    issn = {0894-8755}
}

@article{Roberts2020SensitivityChanges,
    title = {{Sensitivity of the Atlantic Meridional Overturning Circulation to Model Resolution in CMIP6 HighResMIP Simulations and Implications for Future Changes}},
    year = {2020},
    journal = {Journal of Advances in Modeling Earth Systems},
    author = {Roberts, Malcolm J. and Jackson, Laura C. and Roberts, Christopher D. and Meccia, Virna and Docquier, David and Koenigk, Torben and Ortega, Pablo and Moreno-Chamarro, Eduardo and Bellucci, Alessio and Coward, Andrew and Drijfhout, Sybren and Exarchou, Eleftheria and Gutjahr, Oliver and Hewitt, Helene and Iovino, Doroteaciro and Lohmann, Katja and Putrasahan, Dian and Schiemann, Reinhard and Seddon, Jon and Terray, Laurent and Xu, Xiaobiao and Zhang, Qiuying and Chang, Ping and Yeager, Stephen G. and Castruccio, Frederic S. and Zhang, Shaoqing and Wu, Lixin},
    number = {8},
    month = {8},
    pages = {e2019MS002014},
    volume = {12},
    publisher = {John Wiley {\&} Sons, Ltd},
    url = {https://onlinelibrary.wiley.com/doi/full/10.1029/2019MS002014 https://onlinelibrary.wiley.com/doi/abs/10.1029/2019MS002014 https://agupubs.onlinelibrary.wiley.com/doi/10.1029/2019MS002014},
    doi = {10.1029/2019MS002014},
    issn = {1942-2466}
}

@article{Leutbecher2017StochasticVision,
    title = {{Stochastic representations of model uncertainties at ECMWF: state of the art and future vision}},
    year = {2017},
    journal = {Quarterly Journal of the Royal Meteorological Society},
    author = {Leutbecher, Martin and Lock, Sarah Jane and Ollinaho, Pirkka and Lang, Simon T.K. and Balsamo, Gianpaolo and Bechtold, Peter and Bonavita, Massimo and Christensen, Hannah M. and Diamantakis, Michail and Dutra, Emanuel and English, Stephen and Fisher, Michael and Forbes, Richard M. and Goddard, Jacqueline and Haiden, Thomas and Hogan, Robin J. and Juricke, Stephan and Lawrence, Heather and MacLeod, Dave and Magnusson, Linus and Malardel, Sylvie and Massart, Sebastien and Sandu, Irina and Smolarkiewicz, Piotr K. and Subramanian, Aneesh and Vitart, Frédéric and Wedi, Nils and Weisheimer, Antje},
    number = {707},
    month = {7},
    pages = {2315--2339},
    volume = {143},
    publisher = {John Wiley {\&} Sons, Ltd},
    url = {https://onlinelibrary.wiley.com/doi/full/10.1002/qj.3094 https://onlinelibrary.wiley.com/doi/abs/10.1002/qj.3094 https://rmets.onlinelibrary.wiley.com/doi/10.1002/qj.3094},
    doi = {10.1002/QJ.3094},
    issn = {1477-870X}
}

@article{Bourgeois2022Stratification55S,
    title = {{Stratification constrains future heat and carbon uptake in the Southern Ocean between 30{${}^\circ$}S and 55{${}^\circ$}S}},
    year = {2022},
    journal = {Nature Communications},
    author = {Bourgeois, Timothée and Goris, Nadine and Schwinger, Jörg and Tjiputra, Jerry F.},
    number = {},
    month = {1},
    pages = {},
    volume = {13},
    publisher = {Nature Publishing Group},
    url = {https://www.nature.com/articles/s41467-022-27979-5},
    doi = {10.1038/s41467-022-27979-5},
    issn = {2041-1723},
    pmid = {35039511}
}

@article{Moorman2026TheMelt,
    title = {{The Antarctic coastal ocean heat budget is dominated by heat loss to land ice melt}},
    year = {2026},
    journal = {Science Advances},
    author = {Moorman, Ruth and Thompson, Andrew F. and Youngs, Madeleine K. and Stewart, Andrew L.},
    number = {9},
    month = {2},
    volume = {12},
    doi = {10.1126/sciadv.aec7443},
    issn = {2375-2548}
}

@article{Doos1994TheOcean,
    title = {{The Deacon Cell and the Other Meridional Cells of the Southern Ocean}},
    year = {1994},
    journal = {Journal of Physical Oceanography},
    author = {D{\"{o}}{\"{o}}s, Kristofer and Webb, David J.},
    number = {2},
    month = {2},
    pages = {429--442},
    volume = {24},
    publisher = {American Meteorological Society},
    url = {http://journals.ametsoc.org/doi/10.1175/1520-0485(1994)024<0429:TDCATO>2.0.CO;2},
    doi = {10.1175/1520-0485(1994)024<0429:TDCATO>2.0.CO;2},
    issn = {0022-3670}
}

@article{Marshall2017TheStudy,
    title = {{The dependence of the ocean’s MOC on mesoscale eddy diffusivities: A model study}},
    year = {2017},
    journal = {Ocean Modelling},
    author = {Marshall, John and Scott, Jeffery R. and Romanou, Anastasia and Kelley, Maxwell and Leboissetier, Anthony},
    month = {3},
    pages = {1--8},
    volume = {111},
    publisher = {Elsevier},
    doi = {10.1016/J.OCEMOD.2017.01.001},
    issn = {1463-5003}
}

@article{Kim2016TheNeon,
    title = {{The distribution of glacial meltwater in the Amundsen Sea, Antarctica, revealed by dissolved helium and neon}},
    year = {2016},
    journal = {Journal of Geophysical Research: Oceans},
    author = {Kim, Intae and Hahm, Doshik and Rhee, Tae Siek and Kim, Tae Wan and Kim, Chang Sin and Lee, Sang Hoon},
    number = {3},
    month = {3},
    pages = {1654--1666},
    volume = {121},
    publisher = {John Wiley {\&} Sons, Ltd},
    url = {https://agupubs.onlinelibrary.wiley.com/doi/10.1002/2015JC011211},
    doi = {10.1002/2015JC011211},
    issn = {2169-9291}
}

@article{Kopp2010TheExperiments,
    title = {{The impact of Greenland melt on local sea levels: A partially coupled analysis of dynamic and static equilibrium effects in idealized water-hosing experiments}},
    year = {2010},
    journal = {Climatic Change},
    author = {Kopp, Robert E. and Mitrovica, Jerry X. and Griffies, Stephen M. and Yin, Jianjun and Hay, Carling C. and Stouffer, Ronald J.},
    number = {},
    month = {12},
    pages = {619--625},
    volume = {103},
    publisher = {Springer},
    url = {https://link.springer.com/article/10.1007/s10584-010-9935-1},
    doi = {10.1007/s10584-010-9935-1},
    issn = {01650009}
}

@article{Ragen2022TheCirculation,
    title = {{The Role of Atlantic Basin Geometry in Meridional Overturning Circulation}},
    year = {2022},
    journal = {Journal of Physical Oceanography},
    author = {Ragen, Sarah and Armour, Kyle C. and Thompson, LuAnne and Shao, Andrew and Darr, David},
    number = {3},
    month = {3},
    pages = {475--492},
    volume = {52},
    url = {https://journals.ametsoc.org/view/journals/phoc/52/3/JPO-D-21-0036.1.xml},
    doi = {10.1175/JPO-D-21-0036.1},
    issn = {0022-3670}
}

@article{Ehlert2017TheMixing,
    title = {{The Sensitivity of the Proportionality between Temperature Change and Cumulative CO2 Emissions to Ocean Mixing}},
    year = {2017},
    journal = {Journal of Climate},
    author = {Ehlert, Dana and Zickfeld, Kirsten and Eby, Michael and Gillett, Nathan},
    number = {8},
    month = {4},
    pages = {2921--2935},
    volume = {30},
    publisher = {American Meteorological Society},
    url = {https://journals.ametsoc.org/view/journals/clim/30/8/jcli-d-16-0247.1.xml},
    doi = {10.1175/JCLI-D-16-0247.1},
    issn = {0894-8755}
}

@article{Eisenman2024TheFluxes,
    title = {{The Sensitivity of the Spatial Pattern of Sea Level Changes to the Depth of Antarctic Meltwater Fluxes}},
    year = {2024},
    journal = {Geophysical Research Letters},
    author = {Eisenman, Ian and Basinski‐Ferris, Aurora and Beer, Emma and Zanna, Laure},
    number = {19},
    month = {10},
    pages = {e2024GL110633},
    volume = {51},
    url = {https://agupubs.onlinelibrary.wiley.com/doi/10.1029/2024GL110633},
    doi = {10.1029/2024GL110633},
    issn = {0094-8276}
}

@article{Swart2023TheDesign,
    title = {{The Southern Ocean Freshwater Input from Antarctica (SOFIA) Initiative: scientific objectives and experimental design}},
    year = {2023},
    journal = {Geoscientific Model Development},
    author = {Swart, Neil C. and Martin, Torge and Beadling, Rebecca and Chen, Jia-Jia and Danek, Christopher and England, Matthew H. and Farneti, Riccardo and Griffies, Stephen M. and Hattermann, Tore and Hauck, Judith and Haumann, F. Alexander and J{\"{u}}ling, André and Li, Qian and Marshall, John and Muilwijk, Morven and Pauling, Andrew G. and Purich, Ariaan and Smith, Inga J. and Thomas, Max},
    number = {24},
    month = {12},
    pages = {7289--7309},
    volume = {16},
    doi = {10.5194/gmd-16-7289-2023},
    issn = {1991-9603}
}

@article{Liu2023TheSalinity,
    title = {{The Spread of Ocean Heat Uptake Efficiency Traced to Ocean Salinity}},
    year = {2023},
    journal = {Geophysical Research Letters},
    author = {Liu, Maofeng and Soden, Brian J. and Vecchi, Gabriel A. and Wang, Chenggong},
    number = {4},
    month = {2},
    pages = {e2022GL100171},
    volume = {50},
    publisher = {John Wiley {\&} Sons, Ltd},
    url = {https://onlinelibrary.wiley.com/doi/full/10.1029/2022GL100171 https://onlinelibrary.wiley.com/doi/abs/10.1029/2022GL100171 https://agupubs.onlinelibrary.wiley.com/doi/10.1029/2022GL100171},
    doi = {10.1029/2022GL100171},
    issn = {1944-8007}
}

@article{Jones2011TheWinds,
    title = {{The transient response of the Southern Ocean pycnocline to changing atmospheric winds}},
    year = {2011},
    journal = {Geophysical Research Letters},
    author = {Jones, Daniel C. and Ito, Takamitsu and Lovenduski, Nicole S.},
    number = {15},
    month = {8},
    volume = {38},
    publisher = {Blackwell Publishing Ltd},
    url = {https://agupubs.onlinelibrary.wiley.com/doi/full/10.1029/2011GL048145 https://agupubs.onlinelibrary.wiley.com/doi/abs/10.1029/2011GL048145 https://agupubs.onlinelibrary.wiley.com/doi/10.1029/2011GL048145},
    doi = {10.1029/2011GL048145},
    issn = {00948276}
}

@article{MacDougall2017TheParameters,
    title = {{The Uncertainty in the Transient Climate Response to Cumulative CO2 Emissions Arising from the Uncertainty in Physical Climate Parameters}},
    year = {2017},
    journal = {Journal of Climate},
    author = {MacDougall, Andrew H. and Swart, Neil C. and Knutti, Reto},
    number = {2},
    month = {1},
    pages = {813--827},
    volume = {30},
    publisher = {American Meteorological Society},
    url = {https://journals.ametsoc.org/view/journals/clim/30/2/jcli-d-16-0205.1.xml},
    doi = {10.1175/JCLI-D-16-0205.1},
    issn = {0894-8755}
}

@article{Moorman2020ThermalModel,
    title = {{Thermal Responses to Antarctic Ice Shelf Melt in an Eddy-Rich Global Ocean–Sea Ice Model}},
    year = {2020},
    journal = {Journal of Climate},
    author = {Moorman, Ruth and Morrison, Adele K. and Hogg, Andrew McC.},
    number = {15},
    month = {8},
    pages = {6599--6620},
    volume = {33},
    publisher = {American Meteorological Society},
    url = {https://journals.ametsoc.org/view/journals/clim/33/15/jcliD190846.xml},
    doi = {10.1175/JCLI-D-19-0846.1},
    issn = {0894-8755}
}

@article{Sun2020TransientBasins,
    title = {{Transient Overturning Compensation between Atlantic and Indo-Pacific Basins}},
    year = {2020},
    journal = {Journal of Physical Oceanography},
    author = {Sun, Shantong and Thompson, Andrew F. and Eisenman, Ian},
    number = {8},
    month = {7},
    pages = {2151--2172},
    volume = {50},
    publisher = {American Meteorological Society},
    doi = {10.1175/JPO-D-20-0060.1},
    issn = {0022-3670}
}

@article{Jansen2018TransientWarming,
    title = {{Transient versus Equilibrium Response of the Ocean’s Overturning Circulation to Warming}},
    year = {2018},
    journal = {Journal of Climate},
    author = {Jansen, Malte F. and Nadeau, Louis Philippe and Merlis, Timothy M.},
    number = {13},
    month = {7},
    pages = {5147--5163},
    volume = {31},
    publisher = {American Meteorological Society},
    url = {https://journals.ametsoc.org/view/journals/clim/31/13/jcli-d-17-0797.1.xml},
    doi = {10.1175/JCLI-D-17-0797.1},
    issn = {0894-8755}
}

@article{Naughten2023UnavoidableCentury,
    title = {{Unavoidable future increase in West Antarctic ice-shelf melting over the twenty-first century}},
    year = {2023},
    journal = {Nature Climate Change},
    author = {Naughten, Kaitlin A. and Holland, Paul R. and De Rydt, Jan},
    number = {},
    month = {10},
    pages = {1222--1228},
    volume = {13},
    publisher = {Nature Publishing Group},
    url = {https://www.nature.com/articles/s41558-023-01818-x},
    doi = {10.1038/s41558-023-01818-x},
    issn = {1758-6798}
}

@article{Zanna2019UncertaintyPredictions,
    title = {{Uncertainty and scale interactions in ocean ensembles: From seasonal forecasts to multidecadal climate predictions}},
    year = {2019},
    journal = {Quarterly Journal of the Royal Meteorological Society},
    author = {Zanna, L. and Brankart, J. M. and Huber, M. and Leroux, S. and Penduff, T. and Williams, P. D.},
    number = {S1},
    pages = {160--175},
    volume = {145},
    publisher = {John Wiley {\&} Sons, Ltd},
    url = {https://onlinelibrary.wiley.com/doi/full/10.1002/qj.3397 https://onlinelibrary.wiley.com/doi/abs/10.1002/qj.3397 https://rmets.onlinelibrary.wiley.com/doi/10.1002/qj.3397},
    doi = {10.1002/QJ.3397},
    issn = {1477-870X}
}

@article{Garabato2017VigorousShelf,
    title = {{Vigorous lateral export of the meltwater outflow from beneath an Antarctic ice shelf}},
    year = {2017},
    journal = {Nature},
    author = {Garabato, Alberto C. Naveira and Forryan, Alexander and Dutrieux, Pierre and Brannigan, Liam and Biddle, Louise C. and Heywood, Karen J. and Jenkins, Adrian and Firing, Yvonne L. and Kimura, Satoshi},
    number = {},
    month = {1},
    pages = {219--222},
    volume = {542},
    publisher = {Nature Publishing Group},
    url = {https://www.nature.com/articles/nature20825},
    doi = {10.1038/nature20825},
    issn = {1476-4687},
    pmid = {28135723}
}

\end{document}